\documentclass[aps,prb,twocolumn,nofootinbib,groupedaddress,floatfix,superscriptaddress,pdflatex]{revtex4-2}

\usepackage{graphics}
\usepackage{graphicx}
\usepackage{amssymb}
\usepackage{amsmath}
\usepackage{amsfonts}
\usepackage{color}
\usepackage{setspace}
\usepackage{multirow}
\usepackage{array}
\usepackage[subrefformat=parens]{subcaption}
\usepackage{url}

\def\PATHFIG{./figure}

\begin{document}

\title{Regional chemical potential analysis for material surfaces}

\author{Masahiro Fukuda}
\email{masahiro.fukuda@issp.u-tokyo.ac.jp}
\affiliation{Institute for Solid State Physics, The University of Tokyo, 5-1-5 Kashiwanoha, Kashiwa, Chiba 277-8581, Japan}
\author{Masato Senami}
\affiliation{Department of Micro Engineering, Kyoto University, Kyoto 615-8540, Japan}
\author{Yoshiaki Sugimoto}
\affiliation{Department of Advanced Materials Science, The University of Tokyo, Kashiwa, Chiba 277-8561, Japan}
\author{Taisuke Ozaki}
\affiliation{Institute for Solid State Physics, The University of Tokyo, 5-1-5 Kashiwanoha, Kashiwa, Chiba 277-8581, Japan}

\date{\today}

\begin{abstract}
We propose a local regional chemical potential (RCP) analysis method based on an energy window scheme to quantitatively estimate the selectivity of atomic and molecular adsorption on surfaces, as well as the strength of chemical bonding forces between a probe tip and a surface in atomic force microscopy (AFM) measurements.
In particular, focusing on the local picture of covalent bonding,
we use a simple H$_2$ molecular model to demonstrate a clear relationship between chemical bonding forces and the local RCP. 
Moreover, density functional theory calculations on molecular systems and diamond C(001) surfaces reveal that the local RCP at the surfaces
successfully visualizes electron-donating regions such as dangling bonds and double bonds. 
These results suggest that the local RCP can serve as an effective measure to analyze high-resolution non-contact or near-contact AFM images enhanced by chemical bonding forces.
\end{abstract}

\maketitle
\section{Introduction}
\label{sec:Introduction}

The chemical potential is used as a measure to discuss the transfer of electrons resulting from ionic bonding. 
In particular, in the field of density functional theory (DFT), it is known that when two isolated systems A and B come to an equilibrium state by interaction, electron transfer $\Delta N$ and energy change $\Delta E$ can be expressed using chemical potentials $\mu$ and hardness $\eta$ as follows~\cite{parr1989}.
\begin{align}
\mu \equiv \left( \frac{\partial E}{\partial N} \right)_{v}, \ \ \ 
\eta \equiv \frac{1}{2} \left( \frac{\partial^2 E}{\partial N^2} \right)_{v}, \\
\Delta N = \frac{\mu_B^0 - \mu_A^0}{2(\eta_A + \eta_B)}, \\
\Delta E = -\frac{(\mu_B^0 - \mu_A^0)^2}{4(\eta_A + \eta_B)}, 
\end{align}
where $\mu_A^0$ ($\mu_B^0$) is the chemical potential of isolated system A (B),
and $v$ is the external potential.
However, this argument does not include information such as the spatial distribution of electrons and does not reflect the orientation of molecules or the structure of the crystal surface.
To discuss information on specific spatial regions, Tachibana introduced regional chemical potentials (RCPs)~\cite{Tachibana1992} corresponding to the intrinsic Herring-Nichols work function~\cite{RevModPhys.21.185} and discussed chemical reactions along reaction coordinates based on the regional DFT~\cite{Tachibana1992}.

Let us consider dividing the system into $R$ and $R'$ spatial regions.
The electron transfer occurs from one region to another through the interface situated in-between. 
The regional DFT~\cite{tachibana1999} has revealed that even if the system comes to an equilibrium state by electron transfer,
the RCPs $\mu_R$ and $\mu_{R'}$ are not necessarily equal to each other, nor to the chemical potential of the whole system.
This RCP inequality principle has been investigated 
for the electron transfer in a HeH$^+$ molecule based on the regional DFT~\cite{tachibana1999}. 
The local version of the RCP under a linear response approximation is proposed in Refs.~\cite{Szarek2007,10.1063/1.2973634}
in the context of the bond order analysis.
The spatial distribution analysis of the local version of the RCP is demonstrated for a Pt cluster surface to find reactivity regions for hydrogen adsorption~\cite{10.1063/1.3072369}, and for a carbon nanotube surface to investigate lithium atom adsorption~\cite{10.1063/1.3651182}.

In this paper, we propose a more general RCP analysis method using the electronic state around the Fermi surface.
By applying the analysis method to material surfaces,
we also reveal that even in the case of covalent bonding,
the RCP analysis at the material surface
allows us to discuss quantitatively how selectively atoms and molecules
can be adsorbed on a surface.
By replacing the adsorbed molecule with a probe commonly used in non-contact or near-contact Atomic Force Microscopy (AFM), the ease of molecular adsorption can be interpreted as the strength of the interaction force between the probe tip atom and the material surface.
When both the tip and surface are ionized, ionic interactions can be observed, whereas in the case of a neutral probe, such as a Si tip, chemical bonding forces can be detected~\cite{doi:10.1021/acs.nanolett.9b05280}.
In other words, in terms of AFM experiments, RCP has the potential to evaluate bonding forces between the AFM tip and the material surface.

This paper is organized as follows. 
In Secs.~\ref{sec:Regional_chemical_potential} and \ref{sec:surface_definition}, 
the definitions of local RCP and a material surface are introduced in the framework of regional DFT and relativistic gauge invariant theory of quantum electrodynamics (QED).
In Secs.~\ref{sec:Energy_density} and \ref{sec:Covalent_bonding},
the relation between local RCP and the local picture of covalent bonding is revealed using the electronic stress tensor and the regional energy density.
In Sec.~\ref{sec:Simple_model},
local RCP of a hydrogen atom model is evaluated to grasp a simple interpretation of the local RCP at the material surface.
In addition, a direct relation between the chemical bonding force and the local RCP is revealed using H$_2$ molecular models.
After introducing an approximation method for local RCP of general materials based on DFT in Sec.~\ref{sec:Approx_chemical_potential},
computational details and DFT results of various types of organic molecules and C(001) diamond surfaces are discussed in Secs.~\ref{sec:Computational_details} and \ref{sec:DFT_results}.
The last section is devoted to the conclusions and perspectives.

\section{Theory}
\label{sec:Theory}


\subsection{Regional chemical potential}
\label{sec:Regional_chemical_potential}

In the regional DFT, the RCP in region $R$ is defined under the fixed external potential $v$ as
\begin{align}
    \mu_R &\equiv \left(\frac{\partial E_R}{\partial N_R} \right)_{ N_{R' (\ne R)}}, \label{eq:regional_chempot}
\end{align}
where $E_{R}$ and $N_R$ are the total energy and the number of electron in region $R$.
This represents the change in the energy $E_R$ in region $R$ when the number of electrons $N_R$ is varied, while keeping the number of electrons $N_{R'}$ unchanged.
The ordinary chemical potential of the total system in the electronic equilibrium state is represented as follows~\cite{Tachibana1992,tachibana1999,Tachibana2001}.
\begin{align}
\mu &= \mu_R + \sum_{R' \ne R} \left( \frac{\partial E_{R'}}{\partial N_R} \right)_{N_{R'(\ne R)}},\label{eq:chempot}
\end{align}
where the total energy of the system $E=E_R+\displaystyle\sum_{R'} E_{R'(\ne R)}$, and the total number of electron $N = N_R + \displaystyle\sum_{R'} N_{R'(\ne R)}$.
The presence of the second term, which represents the quantum interference effect,
prohibits the equality in between $\mu$ and $\mu_R$~\cite{Tachibana1992}.  
Suppose that the interference term can be ignored,
it satisfies Sanderson’s principle of electronegativity equalization~\cite{doi:10.1126/science.114.2973.670}, $\mu = \mu_R = \mu_{R'} = \cdots$. 

\begin{figure*}
    \centering
    \includegraphics[width=0.7\linewidth]{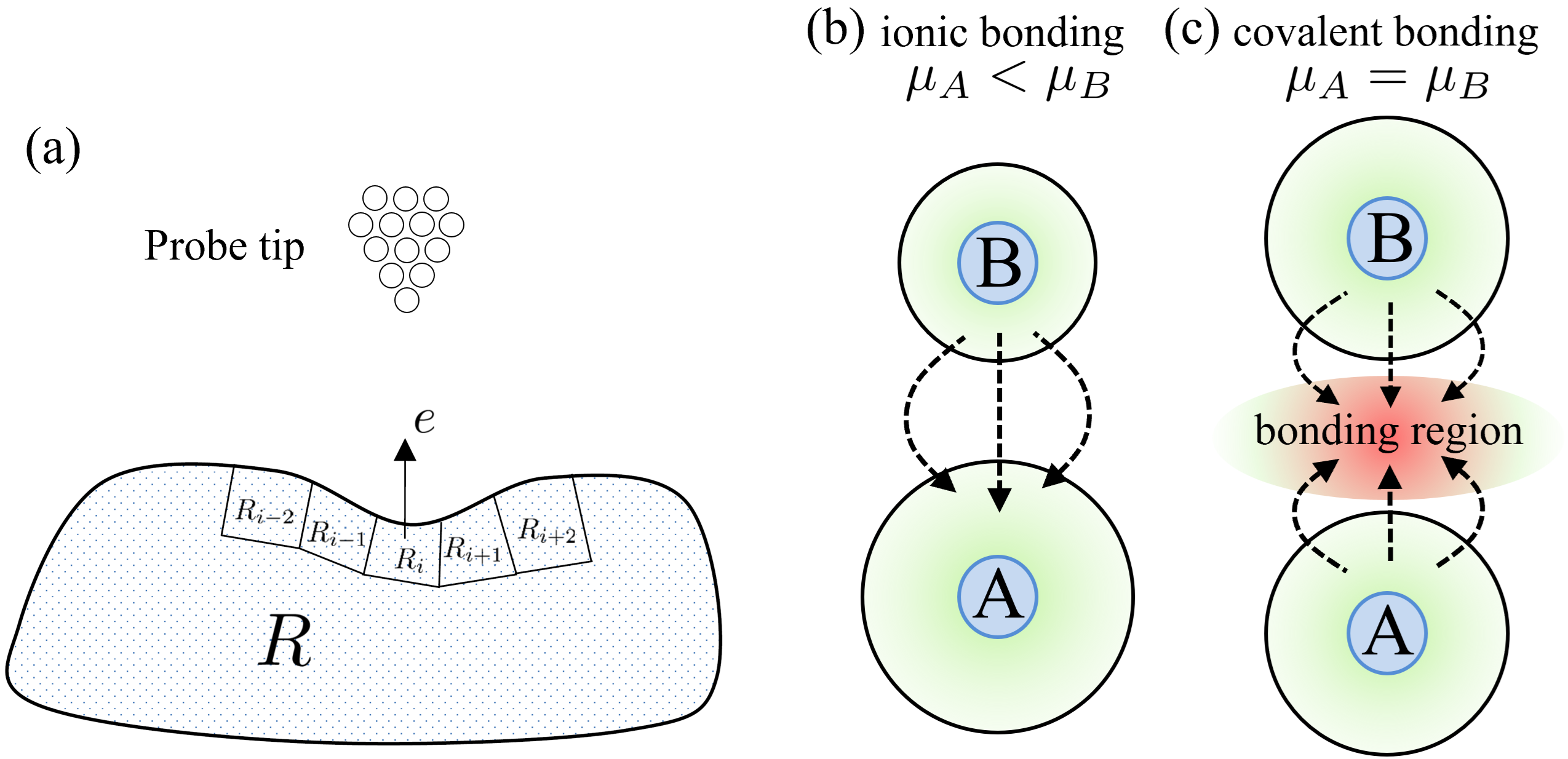}
    \caption{(a) Electron emission from the surface of region $R$.
    (b) Image of ionic bonding and (c) covalent bonding. The circle A (B) represents an atom of the material surface (the apex of the probe tip).}
    \label{fig:electron_emission}
\end{figure*}

To understand the physical meaning of $\mu_R$, let us consider the three-step processes of electron emission shown in Fig.~\ref{fig:electron_emission} (a).
Here, we consider the situation where an electron is emitted from a subdivided region $R_i$ , obtained by dividing the region $R$ into several parts.
(i) Electrons are emitted from the surface of the region $R_i$ to the hypothetical region outside $R$.
(ii) After electrons are emitted from the surface, the potential generated from electrons in the region $R_i$ changes.
(iii) The above change in potential causes electrons in region $R_{j\ne i}$ to flow into region $R_i$ until $\mu$ in region $R$ becomes constant.
From Eq.~\eqref{eq:regional_chempot}, the RCP includes the excitation effects of electrons associated with the processes (i) and (ii),
but note that the process (iii), which corresponds to electron redistribution, is not included.


Since the RCP in Eq.~\eqref{eq:regional_chempot} cannot be calculated exactly in DFT unless the region $R$ is isolated from the other region,
at first, we introduce the finite derivative expression of the RCP as
\begin{align}
    \mu_R \approx \left(\frac{\Delta E_R}{\Delta N_R} \right)_{ N_{R' (\ne R)}}. 
\end{align}
When taking the limit in which the region $R$ becomes infinitesimally small,
the local version of the RCP $\mu_R (\vec{r})$ at each point in space can be expressed as
\begin{align}
    \mu_R (\vec{r}) \approx \frac{\Delta \varepsilon(\vec{r}) }{\Delta n(\vec{r})}, 
\end{align}
where $\Delta \varepsilon(\vec{r})$ and $\Delta n(\vec{r})$ are variations of energy density and electron number density, respectively, under a given perturbative electron excitation.

More detailed approximation expression will be given in Sec.~\ref{sec:Approx_chemical_potential}.
Similar to the conventional chemical potential, which is known to describe the electron transferability,
the local version of the RCP at the surface represents
how much electrons are emitted from or adsorbed onto the material surface.
That will be discussed in Sec.~\ref{sec:Covalent_bonding}.
Before starting the discussion on $\mu_R(\vec{r})$ itself, 
we introduce the definitions of a material surface and the energy density,
which are significant to evaluate $\mu_R(\vec{r})$ on the surface.

\subsection{Definition of a material surface}
\label{sec:surface_definition}

In our scheme, we use the kinetic energy density to define a material surface.
The kinetic energy density is defined in the relativistic gauge invariant theory of QED for electron field and nucleus field (Rigged QED~\cite{Tachibana2001}),
since the U(1) gauge invariance is essential to discuss such local physical quantities without being affected by the phase transformation.
The kinetic energy density operator for electron is defined as~\cite{Tachibana2001}
{\small
\begin{align}
&\hat{T}_e (x) \equiv -\frac{\hbar^2}{4m_e} \sum_{k=1}^3\left[ \hat{\psi}^\dagger (x) \hat{D}_{ek} (x) \hat{D}_{ek} (x) \hat{\psi}(x) \right] + h.c. \\
&\ \xrightarrow{\text{PRQED}} -\frac{\hbar^2}{4m_e} \sum_{k=1}^3\left[ \hat{\psi}^{L\dagger} (x) \hat{D}_{ek} (x) \hat{D}_{ek} (x) \hat{\psi}^L(x) \right] + h.c. , \label{eq:KED}
\end{align}
}%
where $\hat{\psi}(x)$ is the 4-component Dirac field operator, $x=(ct, \vec{r})$, and $c$ is the speed of light in vacuum.
The 4-component $\hat{\psi}(x)$ is decomposed into the large 2-component $\hat{\psi}^L(x)$ and the small 2-component $\hat{\psi}^S(x)$ 
in the Dirac representation as $\hat{\psi}(x) = 
\left(\begin{smallmatrix}
\hat{\psi}^L(x) \\
\hat{\psi}^S(x)
\end{smallmatrix} \right)$.
The gauge covariant derivative is defined as $\hat{D}_{ek}(x)=\partial_k + i \frac{Z_e e}{\hbar c} \hat{A}_k(x)$, where $Z_e =-1$,
$e\, (>0)$ is the elementary charge, $m_e$ is the electron mass, and $\hat{A}_k(x)$ is the vector potential operator.
As we treat non-relativistic systems,
we approximated the expressions of the first equation in the framework of the primary Rigged QED (PRQED) approximation~\cite{Tachibana2001}.
Under the approximation, $\hat{\psi}^S(x) \approx \frac{1}{2m_e c} i \hbar \sigma^k \hat{D}_{ek}(x) \hat{\psi}^L(x)$, where $\sigma^k$ are the $2\times2$ Pauli matrices,
and the spin-dependent terms are ignored after the 2-component decomposition. 

Although the kinetic energy is always positive in classical mechanics,
the value of $\langle \hat{T}_e(x) \rangle$ in quantum theory is not always positive.
The meaning of the sign of $\langle \hat{T}_e(x) \rangle$ can be explained as follows.
For example, when we assume that electrons are trapped inside a potential well,
$\langle \hat{T}_e(x) \rangle > 0$ in the potential well, while $\langle \hat{T}_e(x) \rangle < 0$ outside the well.
While the region where $\langle \hat{T}_e(x) \rangle > 0$ can be understood as the electron is allowed to move from the classical mechanical view,
the region where $\langle \hat{T}_e(x) \rangle < 0$ can be understood as a tunneling region from the quantum theoretical view.
Therefore, the boundary $\langle \hat{T}_e(x) \rangle = 0$ represents the surface of the material.
This definition of the material surface is useful,
since the zero kinetic energy density is determined by the solution of the electronic state
without introducing any artificial or arbitrary parameters.
The zero kinetic energy density isosurface is called ``electronic interface''~\cite{Tachibana2001,https://doi.org/10.1002/qua.25073}. 
For atoms and molecules, there are some electronic interfaces inside of inner electronic structures,
since they have a shell structure.
However, it is easy to imagine that the outermost electronic interface defines the shape of atoms and molecules.
The shape of the electronic interface of atoms is investigated in detail in Ref.~\cite{https://doi.org/10.1002/qua.25073}.

\subsection{Energy density and electronic stress tensor}
\label{sec:Energy_density}
The energy density $\varepsilon (\vec{r})$ is the 0-0 component of the symmetric energy momentum tensor density $T^{\mu \nu}(\vec{r})$,
which is defined in the general relativistic QED theory~\cite{weinberg1972gravitation,Tachibana2012,Tachibana2014,Fukuda2016_LPQ,fukuda2016_spinvorticity,Tachibana2017}.
Suppose that the system is in an equilibrium stationary state in the Minkowski space limit.
In that case, the virial theorem can be applied.
Furthermore, under the Born-Oppenheimer approximation, the total energy of the Rigged QED system is given as~\cite{Tachibana2004,Tachibana2017} 
\begin{align}
    E_{\rm Rigged\ QED} &= \int \varepsilon (\vec{r}) d^3 \vec{r}, \label{eq:energy_density}\\
    &\approx \int \langle m_e c^2 \hat{\psi}^\dagger(\vec{r}) \gamma^0 \hat{\psi}(\vec{r}) \rangle d^3 \vec{r},
\end{align}
where $\gamma^0$ is the $4\times 4$ gamma matrix.
Under the PRQED approximation for taking the non-relativistic limit,  
the total energy minus the mass energy is reduced to
\begin{align}
    E_{\rm PRQED} \approx \int \frac{1}{2} \sum_{k=1}^3 \langle \hat{\tau}_e^{Skk}(\vec{r}) \rangle d^3 \vec{r}. \label{eq:int_taue}
\end{align}
The detailed derivation is given in Sec.~S2 in the Supplemental Information(SI).
The symmetric electronic stress tensor operator in the PRQED is defined as
{\small
\begin{align}
\hat{\tau}_e^{Skl} (x) &\equiv - \hat{T}_e^{kl}(x), \\
&\approx \frac{\hbar^2}{4m_e} \left[ \hat{\psi}^{L \dagger} \hat{D}_{ek} \hat{D}_{el} \hat{\psi}^L - \left( \hat{D}_{ek} \hat{\psi}^L \right)^{\dagger} \hat{D}_{el} \hat{\psi}^L \right] + h.c., \label{eq:stress}
\end{align}
}%
where $\hat{T}_e^{kl}(x)$ is the electronic symmetric energy momentum tensor density operator\footnote{
The scalar expression $\hat{T}_e$ denotes the kinetic energy density operator,
while the second-rank tensor $\hat{T}_e^{kl}$ denotes the electronic symmetric energy momentum tensor operator.
}.
The expression of Eq.~\eqref{eq:int_taue} reveals that only the integral of the trace of the electronic stress tensor
is needed to evaluate the total energy under the above approximations.
Using Eq.~\eqref{eq:int_taue}, we introduce another energy density operator expression as 
\begin{align}
    \varepsilon_\tau(\vec{r})  \equiv \frac{1}{2} \sum_{k=1}^3 \langle \hat{\tau}_e^{Skk}(\vec{r}) \rangle d^3 \vec{r}. \label{eq:E_tau}
\end{align}
Here, $\varepsilon_\tau(\vec{r})$ is referred to as the ``regional energy density''\cite{Tachibana2001,Szarek2007,Tachibana2017}.
Its definition differs from that of the non-relativistic version of the original energy density $\varepsilon(\vec{r})$ in Eq.~\eqref{eq:energy_density}, even when excluding the mass energy density.
However, the energy obtained by integrating over the whole space excluding the mass energy is identical.
The advantage of this regional energy density lies in the fact that
$\varepsilon_\tau(\vec{r})$ can be derived directly from the electronic density matrices, without the need for additional intricate calculations of the interaction energies.
Furthermore, the electronic stress tensor density appeared in $\varepsilon_\tau(\vec{r})$
is known to be significant for a discussion of chemical bonding theory
as mentioned in the next section.  

\subsection{Covalent bonding}
\label{sec:Covalent_bonding}
Let us consider a system such as near-contact AFM measurements~\cite{Zhang2025,PhysRevResearch.7.023036,Moll_2010,doi:10.1021/acs.nanolett.9b05280}, in which the interaction between the material surface and the probe tip gives rise to forces such as ionic bonding, covalent bonding, van der Waals interactions.
For ionic systems, it is easy to understand that the difference in the RCP between the material surface and the probe tip is a measure for the ionic bonding,
since electrons tend to move from regions of higher chemical potential to regions of lower chemical potential, as shown in Fig.~\ref{fig:electron_emission} (b).
What can be said about covalent bonding?
Even if the ordinary chemical potentials are the same over the system, the RCPs can be different in each region.
In such a case, it can be assumed that electrons in the highly electron-donating region,
the higher $\mu_R(\vec{r})$ region, are more likely to be supplied to the bonding region and contribute to covalent bonding as shown in Fig.~\ref{fig:electron_emission} (c). 

Next, we relate this physical interpretation to a local mechanical picture of covalent bonding~\cite{Tachibana2005}.
Normally, the eigenvalues of the electron stress tensor density are negative
because the electrons repel each other and try to diffuse.
However, in a covalent bonding state, the electrons in the bonding region
are in a tensile stress state;
namely the inphase overlap of orbitals brings positive maximum eigenvalues of $\langle \hat{\tau}_e^{Skl} (\vec{r}) \rangle$.
This is known as a local mechanical picture of covalent bonding~\cite{Tachibana2004} (see Fig.~S1 in SI).
In addition, the regional energy density $\varepsilon_\tau(\vec{r}\,)$,
which is a half of the trace of the stress tensor density $\langle \hat{\tau}_e^{Skl} (\vec{r}\,) \rangle$,
increases around the bonding region in the tensile stress state.
As can be seen from the calculations for a hydrogen molecule,
covalent bonding increases the regional energy density in the bonding region
and decreases the regional energy density near the nuclei (see Fig.~S2 in SI).
The total energy as an integral over the whole space is then reduced, 
and the system is stabilized.
This is a mechanism of the covalent bonding stabilization in terms of the regional energy density.
Motivating this chemical bonding picture,
we define the local RCP $\mu_R^{\tau} (\vec{r})$ 
and its inverse, ``local density response function'' $\xi_R^{\tau} (\vec{r})$, using the regional energy density as
\begin{align}
    \mu_R^{\tau} (\vec{r}) = \frac{\Delta \varepsilon_{\tau}(\vec{r}) }{\Delta n(\vec{r})},\ \ \ 
    \xi_R^{\tau} (\vec{r}) = \frac{\Delta n(\vec{r})}{\Delta \varepsilon_{\tau}(\vec{r}) },
\end{align}
where $\xi^\tau_R(\vec{r}\,)$ represents how much electrons are transferred by the energy change. 
If $\xi^\tau_R(\vec{r}\,)$ is negative and has a large absolute value, 
it means that electrons are more likely to transfer from the region $R$
when the energy increases in the bonding region due to tensile stress
during the covalent bond formation process. 
The stronger interaction facilitates the transfer of electrons from region $R$.
That increases the electron overlap, and it makes the interaction stronger.
Although both $\mu^\tau_R(\vec{r}\,)$ and $\xi^\tau_R(\vec{r}\,)$ might be useful for analysis of electron transfer and chemical bonding, 
$\mu^\tau_R(\vec{r}\,)$ is adopted in this study.

\section{Simple model}
\label{sec:Simple_model}

In this section, we first calculate the local RCP $\mu_R^{\tau}(\vec{r})$ on a material surface using the spherical $s$-type Slater-type orbital (STO) as a hydrogen atom model.
The STO is represented in the spherical coordinate as
\begin{align}
\psi^{\rm STO}(r) &= \sqrt{n (r_s)} e^{- \alpha (r-r_s)} = A e^{- \alpha r},
\end{align}
where $A=\sqrt{n (r_s)} e^{\alpha r_s}$. The parameter $r_s$ is the distance between the nucleus position ($r=0$) and the electronic interface,
which is the material surface defined by the isosurface of the zero kinetic energy density as discussed in Sec.~\ref{sec:surface_definition}.
Substituting $\psi^L = 
\left(\begin{smallmatrix}
\psi^{\rm STO}(r)\\
0
\end{smallmatrix} \right)
$ into Eqs.~\eqref{eq:KED}, \eqref{eq:stress}, and \eqref{eq:E_tau},
the kinetic energy density $T_e(r)$ and the regional energy density $\varepsilon_\tau (r)$ are represented as
\begin{align}
T_e(r) &= -\frac{\hbar^2}{2m} \left( \alpha^2 - \alpha \frac{2}{r} \right) A^2 e^{-2 \alpha r}, \\
\varepsilon_\tau (r)
&= - \frac{\hbar^2}{2m} \alpha \frac{1}{r} A^2 e^{-2 \alpha r} .
\end{align}
The RCP at the electronic interface is given by
\begin{align}
\mu_R^\tau (r_s) 
&= \frac{\Delta \varepsilon_\tau (r_s)}{\Delta \rho(r_s)} 
\approx \frac{\varepsilon_\tau (r_s)}{\rho(r_s)} 
= - \frac{\hbar^2}{2m} \alpha \frac{1}{r_s}. \label{eq:mu_s_orbital}
\end{align}
The above linear response approximation is reasonable, since we assumed that this system has only one electron.
From Eq.~\eqref{eq:mu_s_orbital},
we can find $\mu_R^{\tau}$ at the surface is directly related to the decay constant $\alpha$ of the $s$-type STO. 
It can be a measure of delocalization of electrons from near the surface outward.
By calculating the values of $r_s$ and $\mu_R^\tau (r_s)$ from a DFT result of a hydrogen atom, the value of $\alpha$ can be estimated from Eq.~\eqref{eq:mu_s_orbital}.
For the hydrogen atom, the DFT result gives $r_s=2.0$[a.u.], $\mu_R^\tau (r_s)=-0.23$[a.u.] and $\alpha=0.92$.
This value of $\alpha$ is subsequently applied in the force calculation for an H$_2$ molecular model described in the next paragraph.
The computational details of DFT calculations will be explained in a later section.

As a next step, we assume an H$_2$ molecular model to investigate the chemical bonding force between two atomic models separated by distance $l$.
When the $\psi^{\rm STO}$ of each atom is constructed from DFT calculations of an isolated hydrogen atom,
using a linear combination of the spherical STOs, the H$_2$ molecular model can be written as
\begin{align}
    \Psi^{\rm sph}(\vec{r}) &= c_s \psi_s(\vec{r}) + c_t \psi_t(\vec{r}),\\
    \psi_{s}(\vec{r})  &= e^{- \alpha_s |\vec{r}-\vec{r}_s|},\ \ \ 
    \psi_{t}(\vec{r})  = e^{- \beta_t |\vec{r}-\vec{r}_t|},
\end{align}
where $\vec{r}_s = (0,0,0)$ and $\vec{r}_t = (0,0,l)$ in Cartesian coordinate.

\begin{figure*}
    \centering
    \includegraphics[width=0.7\linewidth]{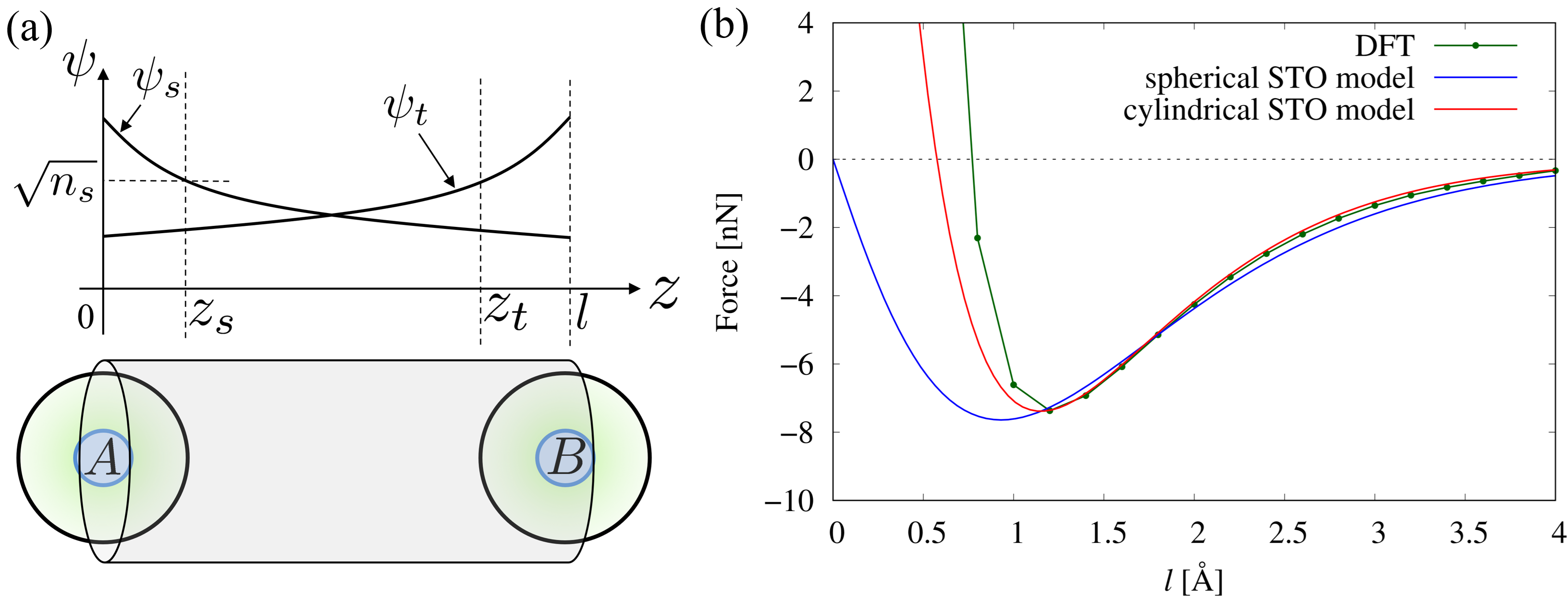}
    \caption{(a) A diatomic molecular model and its wave functions. 
    (b) Force curves of DFT (green), spherical STO model (blue) and cylindrical STO model (red).
    The parameters, ${\cal E}_s \approx \mu_R^\tau(r_s) = -0.23$[a.u.], $\alpha_s=0.92$, $K=1.75$~\cite{10.1063/1.1734456}, and $\sqrt{ \frac{|\Lambda_{st}|^2 }{\Lambda_{ss} \Lambda_{tt}}} \approx 1$ were used.
    } 
    \label{fig:model_H2}
\end{figure*}

We also introduce the simpler model wave function using cylindrical STOs, which can be written as
\begin{align}
    \Psi^{\rm cyl}(z,r,\theta) &= c_s \psi_s(z,r,\theta) + c_t \psi_t(z,r,\theta)  \ \ \ (0<z<l),\\
    \psi_{s}  &= \sqrt{n_s} e^{- \alpha_s (z-z_s)}  \Lambda_s(r,\theta),\\
    \psi_{t}  &= \sqrt{n_t} e^{\beta_t (z-z_t)} \Lambda_t(r,\theta), 
\end{align}
where $n_s$ and $n_t$ are the electron number density at the electronic interface ($z=z_s, z_t$), which is
defined by the zero kinetic energy density of the single atomic model.
This model is considered to be based on the assumption that
the atom $s$ on $z=0$ represents an atom of a material surface and the atom $t$ on $z=l$ represents an apex atom of a tip used in AFM experiments as shown in Fig.~\ref{fig:model_H2} (a).
$\Lambda_s(r,\theta)$ and $\Lambda_t(r,\theta)$ are functions of the polar coordinates.
Later, we will see that this simpler model is enough to reproduce a force curve of H$_2$ molecule.

To obtain the ground state of this model, 
we have to solve $|ES - H| = 0$, where
\begin{align}
H = 
\begin{bmatrix}
H_{ss} & H_{st}\\
H_{ts} & H_{tt}
\end{bmatrix}
,\ \ \ 
S = 
\begin{bmatrix}
S_{ss} & S_{st}\\
S_{ts} & S_{tt}
\end{bmatrix}
,
\end{align}
\begin{align}
H_{ij} &= \int d^3\vec{r} \psi_i^\dagger \hat{H} \psi_j,\\
S_{ij} &= \int d^3\vec{r} \psi_i^\dagger \psi_j.
\end{align}
Under the extended H\"{u}ckel Method~\cite{10.1063/1.1734456},
suppose that $H_{ss} \approx {\cal E}_s S_{ss}$, $H_{tt} \approx {\cal E}_t S_{tt}$, $H_{st} \approx  K S_{st} \frac{{\cal E}_s + {\cal E}_t }{2}$,
and $\tilde{S} \equiv \frac{S_{st} S_{ts}}{S_{ss} S_{tt}} \ll 1$,
we obtain
\begin{align}
E_{\pm} \approx
 \frac{1}{2} \left(
        {\cal E}_{s+t} \pm \sqrt{
                {\cal E}_{s-t}^2  +\left( - {\cal E}_{s-t}^2 + {\cal E}_{s+t}^2 (K-1)^2 \right) \tilde{S}
        } \right),
\end{align}
where ${\cal E}_{s \pm t} \equiv {\cal E}_s \pm {\cal E}_t$.
The negative (positive) sign in front of the square root corresponds to the solution of covalent (anti-covalent) bonding.
Furthermore, we assume that ${\cal E}_s = {\cal E}_t$, $\alpha_s = \beta_t$, and $n_s=n_t$.
The dependency of the distance $l$ is included in $\tilde{S}$ through the overlap matrix $S_{st}$.
Therefore, the two eigenenergies $E_{\pm}$ are approximately given as 
\begin{align}
        E_{\pm} &\approx  {\cal E}_{s}  
        \pm \left|( K-1){\cal E}_s \right| \sqrt{ \tilde{S} }.
\end{align}
For the H$_2$ molecular model, two degenerate electrons occupy the lower energy state. 
Since the Coulomb interaction between the nuclei is negligible
if the distance $l$ is sufficiently large,
the total energy $E_{\rm tot}$ and the chemical bonding force $F(l)$ are given as follows.
\begin{align}
        E_{\rm tot} &= 2E_{-} \approx 2{\cal E}_{s}  
        - \left|2( K-1){\cal E}_s \right| \sqrt{ \tilde{S} }, \\
        F(l) &= -\frac{dE_{\rm tot}}{dl} 
        \approx -|2  (K-1){\cal E}_s| \cdot \frac{d \sqrt{\tilde{S}}}{dl}, \label{eq:force}
\end{align}
where $F(l)$ represents a force between two atoms separated by the distance $l$.

In the case of the spherical STO model $\Psi^{\rm sph}$,
substituting $\tilde{S}= \left(1+\alpha_s l + \frac{1}{3}(\alpha_s l)^2 \right)^2 e^{-2\alpha_s l}$ into Eq.~\eqref{eq:force},
\begin{align}
        F(l) & \approx - \frac{2}{3}|(K-1){\cal E}_s| \cdot (1 + \alpha_s l) \alpha_s^2 l e^{- \alpha_s l}.  \label{eq:force_sph}
\end{align}
In the case of the cylindrical STO model $\Psi^{\rm cyl}$,
substituting $\tilde{S}= \frac{|\Lambda_{st}|^2 }{\Lambda_{ss} \Lambda_{tt}} l^2 e^{-2\alpha_s l}$ into Eq.~\eqref{eq:force},
\begin{align}
        F(l) & \approx  2|(K-1){\cal E}_s| \cdot 2\alpha_s \sqrt{ \frac{|\Lambda_{st}|^2 }{\Lambda_{ss} \Lambda_{tt}}}\left(1 - \alpha_s l \right) e^{- \alpha_s l}. \label{eq:force_cyl}
\end{align}
The overlap integrals for the polar coordinate $\Lambda_{ij}$ are defined as
\begin{align}
    \Lambda_{ij} \equiv \int_0^{r_{\rm cut}} dr \int_0^{2 \pi} d \theta \Lambda_i^*(r,\theta) \Lambda_j(r,\theta),
\end{align}
where $r_{\rm cut}$ is a parameter for the cutoff radius.
Eqs.~\eqref{eq:force_sph} and \eqref{eq:force_cyl} reveal that the shape of the force curve is determined by the value of $\alpha_s$.
Since $\mu_R^{\tau}(r_s)$ gives $\alpha_s$ via Eq.~\eqref{eq:mu_s_orbital},
$\mu_R^{\tau}(\vec{r})$ at electronic interfaces can be a measure of the strength of the chemical bonding force.
This fact supports that a higher value of $\mu_R^{\tau}(\vec{r})$ at the surface suggests a stronger electron-donating ability toward the bonding region, which contributes to covalent bonding, as discussed in the previous section.

A comparison of the above Eqs.~\eqref{eq:force_sph} and \eqref{eq:force_cyl} with a DFT result is shown in Fig.~\ref{fig:model_H2} (b).
The parameters, ${\cal E}_s \approx \mu_R^\tau(r_s) = -0.23$[a.u.], $\alpha_s=0.92$, $K=1.75$~\cite{10.1063/1.1734456}, and $\sqrt{ \frac{|\Lambda_{st}|^2 }{\Lambda_{ss} \Lambda_{tt}}} \approx 1$ were used for plotting Fig.~\ref{fig:model_H2}(b). 
The values of $\mu_R^\tau(r_s)$ and $\alpha_s$ were obtained by the hydrogen atom model. 
Considering that the atomic surfaces are almost in contact at $l=2 r_s=2.11$ \AA,
these models are in very good agreement at the chemical bonding region, $l>2.11$ \AA.

In this section, we have revealed the relation between $\mu_R^\tau(r_s)$ and the force $F(l)$ using a simple H$_2$ model.
However, the approximation in Eq.~\eqref{eq:mu_s_orbital} cannot be applied to general materials.
The extension of the approximation of $\mu_R^\tau(\vec{r})$ in general materials is discussed in the next section.

\section{Approximation for regional chemical potential}
\label{sec:Approx_chemical_potential}

\begin{figure}
    \centering
    \includegraphics[width=0.9\linewidth]{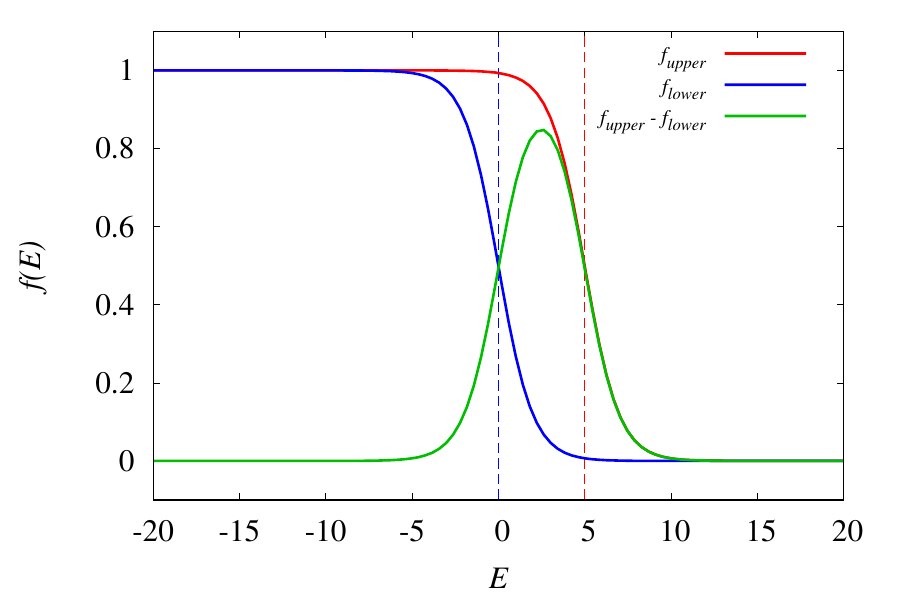}
    \caption{An example of the Fermi functions $f_{\rm upper}$
    and $f_{\rm lower}$.
    The parameters are set as $\mu_{\rm lower} = 0$, $\mu_{\rm upper}=5$, $\tilde{T}_{\rm lower}=\tilde{T}_{\rm upper} =1$.}
    \label{fig:Fermi_function}
\end{figure}

For calculations of the RCP, the response of the energy density to the change of electron number density is needed.
However, it is not easy to obtain such a response function from the DFT calculations.
For actual calculations of the RCP, 
we have to introduce some approximations.
At first, we redefine $\mu_R$ using variations in the local energy density $\Delta \varepsilon_{\tau}(\vec{r})$ and the local electron number density $\Delta n(\vec{r})$ for the global chemical potential change $\Delta \mu$ as follows.
\begin{align}
\mu_R \to \mu_R^{\tau} (\vec{r}) 
\equiv \frac{\Delta \varepsilon_{\tau}(\vec{r}) }{\Delta n(\vec{r})}
\approx \frac{\varepsilon_{\tau, {\rm ew}}(\vec{r}) }{n_{\rm ew}(\vec{r})}.
\end{align}
We call the last approximation an ``energy window scheme",
where the physical quantities are calculated in an energy range of 
$\mu_{\rm lower} < \mu < \mu_{\rm upper}$ using the Fermi function with broadening factors $\tilde{T}_{\rm lower}$ and $\tilde{T}_{\rm upper}$ as follows. 
{\small
\begin{align}
n_{\rm ew}(\vec{r}) 
&= \int_{-\infty}^{\infty} dE n(\vec{r}, E) \left[ f_{\rm upper}(E) - f_{\rm lower} (E)\right], \label{eq:n_ew}\\
\varepsilon_{\tau, \rm ew}(\vec{r})
&= \int_{-\infty}^{\infty} dE \varepsilon_\tau(\vec{r}, E) \left[  f_{\rm upper}(E) - f_{\rm lower} (E) \right], \label{eq:e_ew}
\end{align}
}%
where the upper and lower Fermi functions are defined as
$f_\lambda(E) \equiv f(E, \mu_\lambda, \tilde{T}_\lambda) \equiv 1/(1+ e^{(E-\mu_\lambda)/\tilde{T}_\lambda })$
as shown in Fig.~\ref{fig:Fermi_function}.
The energy dependent electron number density $n(\vec{r},E)$ and regional energy density $\varepsilon_\tau(\vec{r}, E)$ are given in 
Appendix \ref{sec:Supp_Approx_DFT}.
The four parameters $\mu_{\rm lower}$, $\mu_{\rm upper}$, $\tilde{T}_{\rm lower}$, and $\tilde{T}_{\rm upper}$ should be properly selected
for the purpose of physical considerations.

\begin{figure*}
    \centering
    \includegraphics[width=0.8\linewidth]{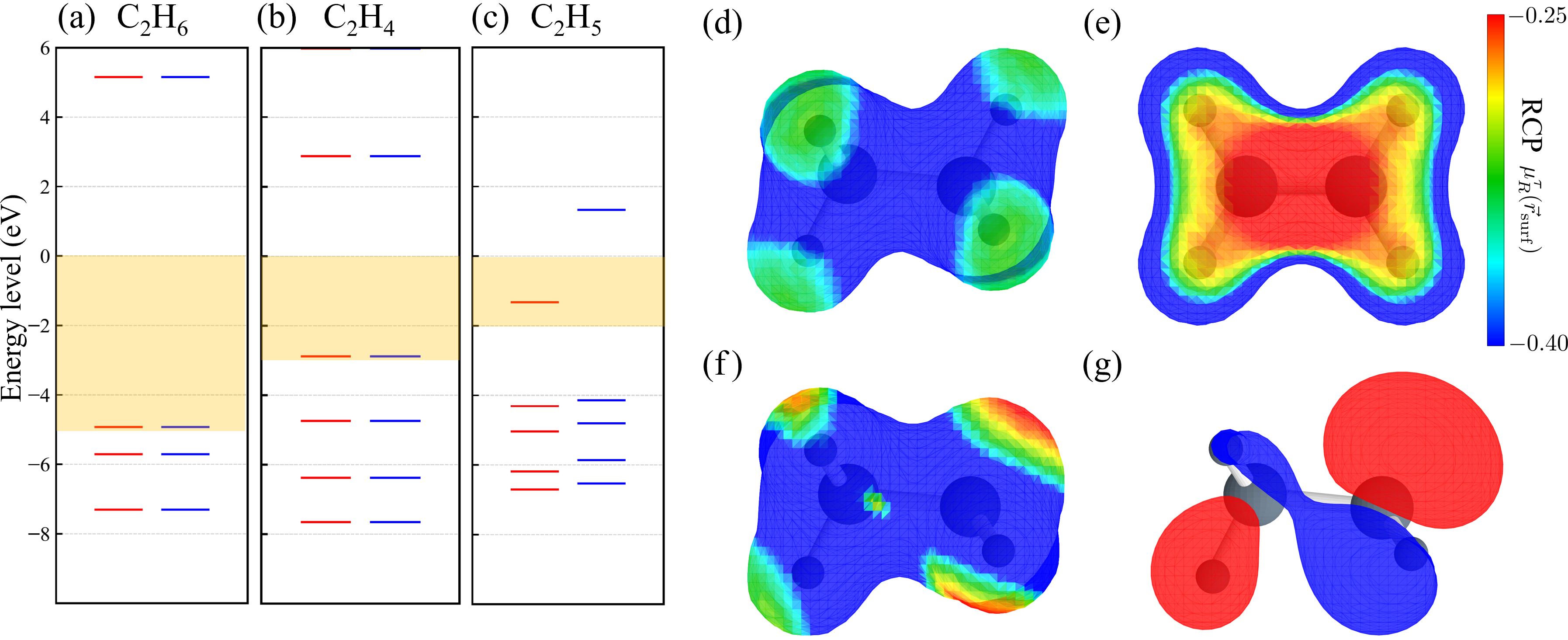}
    \caption{Energy levels of (a) C$_2$H$_6$, (b) C$_2$H$_4$, and (c) C$_2$H$_5$ molecular models. The area colored in orange represents the energy window for local RCP $\mu_R^{\tau}(\vec{r})$.
    The RCP on the surface for (d) C$_2$H$_6$, (e) C$_2$H$_4$, and (f) C$_2$H$_5$ molecular models are drawn as color maps on the surface.
    (g) The HOMO of C$_2$H$_5$ are drawn as the red (+0.05 a.u.) and blue (-0.05 a.u.) isosurfaces.}
    \label{fig:ethane_type}
\end{figure*}
\section{Computational details}
\label{sec:Computational_details}
The first-principles calculations based on DFT were carried out for
H atom, H$_2$ molecule,
some organic molecules, and C(001) diamond surface.
The DFT calculations within a generalized gradient approximation (GGA) \cite{PhysRev.140.A1133,PhysRevLett.77.3865} 
were performed for the geometry relaxations and variable cell optimizations using the
OpenMX code~\cite{OpenMX} based on norm-conserving pseudopotentials generated
with multireference energies \cite{PhysRevB.47.6728} and optimized pseudoatomic
basis functions~\cite{PhysRevB.67.155108}.
The standard basis set we used are listed in OpenMX website~\cite{OpenMX_basisset}.
The qualities of the basis functions and fully relativistic pseudopotentials were carefully
benchmarked by the delta gauge method~\cite{Lejaeghere2016-gh}
to ensure the accuracy of our calculations.
An electronic temperature of 300 K is used to count the number
of electrons by the Fermi-Dirac function.
The regular mesh of 240 Ry in real space was used for the numerical
integration and for the solution of the Poisson equation~\cite{PhysRevB.72.045121,DUY2014777}.
Cell vectors and internal coordinates are optimized by using a combination scheme of
the rational function (RF) method~\cite{doi:10.1021/j100247a015} 
and the direct inversion iterative sub-space (DIIS) method~\cite{CSASZAR198431}
with a BFGS update~\cite{10.1093/imamat/6.1.76, 10.1093/comjnl/13.3.317, 10.2307/2004873, 10.2307/2004840} for the approximate Hessian. 
The force on each atom was relaxed to be less than 0.0003 Hartree/Bohr.
For the C(001) diamond surface calculations, we used the same computational conditions in Ref.~\cite{PhysRevResearch.7.023036}.

Calculations of the local RCPs and the kinetic energy density were performed by 
FLPQ module~\cite{FLPQ} in QEDalpha package~\cite{QEDalpha}. 
FLPQViewer~\cite{FLPQViewer} was used for visualization of the atomic structures, RCPs, and other properties.
Similar functions are available in 3D Viewers such as OpenMX Viewer~\cite{LEE2019192} and Fermisurfer~\cite{KAWAMURA2019197}.
The atomic unit is used for $\mu_R^{\tau}(\vec{r})$ (i.e. 1[Hartree/Bohr$^3$] = 183.631[eV/\AA$^3$]). 
The broadening factors appeared in Eqs.~\eqref{eq:n_ew} and \eqref{eq:e_ew} were set to $\tilde{T}_{\rm lower} = \tilde{T}_{\rm upper} = 0.001 / k_B [{\rm K}]$ for all the systems, where the Boltzmann constant $k_B =8.617 \times 10^{-5} [{\rm eV}/{\rm K}]$.


\section{DFT results}
\label{sec:DFT_results}

\subsection{C$_2$H$_6$, C$_2$H$_4$, and C$_2$H$_5$ molecules}
\label{sec:molecules}
Figures \ref{fig:ethane_type} (a)-(c) show the energy levels of C$_2$H$_6$, C$_2$H$_4$, and C$_2$H$_5$ molecular models, respectively.
To draw figures of each local RCP on the electronic interfaces $\mu_R^{\tau}(\vec{r}_{\rm {surf}})$,
at least HOMO (Highest Occupied Molecular Orbital) should be included in the energy window $\mu_{\rm lower} < \mu < \mu_{\rm upper}$.
Therefore, the energy windows were chosen to be -5\,eV to 0\,eV for C$_2$H$_6$, 
-3\,eV to 0\,eV for C$_2$H$_4$, and -2\,eV to 0\,eV for C$_2$H$_5$.
The corresponding $\mu_R^{\tau}(\vec{r}_{\rm {surf}})$ are shown in
Figs.~\ref{fig:ethane_type} (d)-(f).

In Fig.~\ref{fig:ethane_type} (d), 
although the H atoms in C$_2$H$_6$ are visualized as green parts,
overall $\mu_R^{\tau}(\vec{r}_{\rm surf})$ is small.
In contrast, the C$_2$H$_4$ (Fig.~\ref{fig:ethane_type} (e)) has the high $\mu_R^{\tau}(\vec{r}_{\rm surf})$ (red regions) originating from a C-C double bond.
This result is consistent with the fact that the double bonding region is 
a highly electron-donating region,
where electrons are more likely to be supplied to outer regions. 
Figure \ref{fig:ethane_type} (f) shows a molecule model whose structure is generated 
as one H atom is deleted from C$_2$H$_6$ without geometry optimization. 
The red region at the top right in the C$_2$H$_5$ corresponds to the high $\mu_R^{\tau}(\vec{r}_{\rm surf})$ arising from a dangling bond
that was formerly bonded to the deleted H atom.
This fact is consistent with the HOMO distribution depicted in Fig.~\ref{fig:ethane_type} (g).

\begin{figure*}
    \centering
    \includegraphics[width=0.8\linewidth]{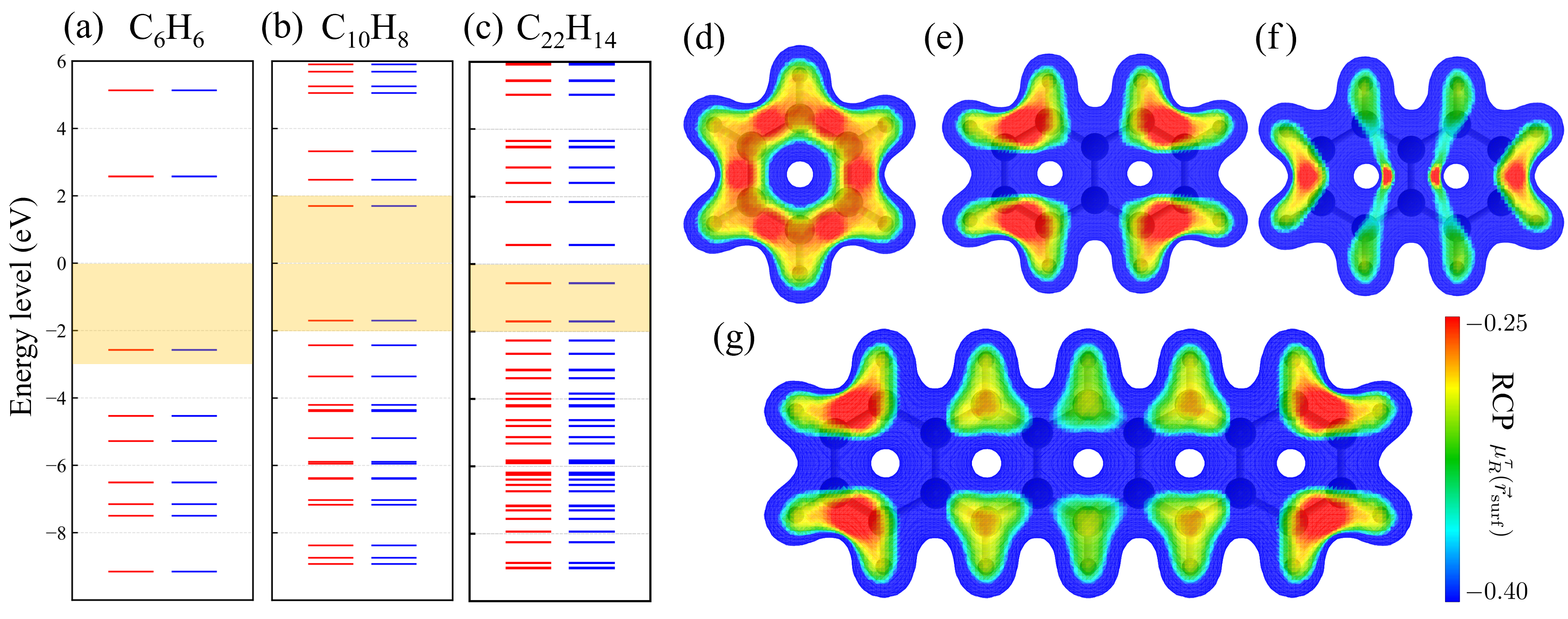}
    \caption{Energy levels of (a) benzene (C$_6$H$_6$), (b) naphthalene (C$_{10}$H$_8$), and (c) pentacene (C$_{22}$H$_{14}$) molecules.
    The area colored in orange represents the energy window for RCP $\mu_R^{\tau}(\vec{r})$.
    The RCP $\mu_R^{\tau}(\vec{r})$ on the surface for (d) benzene (C$_6$H$_6$), (e) naphthalene (C$_{10}$H$_8$) in the energy range between -2\,eV and 0\,eV, (f) naphthalene (C$_{10}$H$_8$) in the energy range between 0\,eV and 2\,eV,
    and (g) pentacene (C$_{22}$H$_{14}$) molecules are drawn as color maps on the surface.}
    \label{fig:benzene_type}
\end{figure*}

\subsection{Benzene, naphthalene, and pentacene molecules}
\label{sec:molecules2}
%
The energy windows employed in the calculation of the RCPs $\mu_R^{\tau}(\vec{r}_{\rm{surf}})$ for benzene (C$_6$H$_6$), naphthalene (C$_{10}$H$_8$), and pentacene (C$_{22}$H$_{14}$) are shown in Figs.~\ref{fig:benzene_type} (a)–(c), and the corresponding RCP distributions are presented in Figs.~\ref{fig:benzene_type} (d)–(g).
For pentacene, RCP $\mu_R^{\tau}(\vec{r}_{\rm {surf}})$ in the energy range between -2\,eV and 0\,eV is shown in Fig.~\ref{fig:benzene_type} (e),
and one in the energy range between 0\,eV and 2\,eV is shown in Fig.~\ref{fig:benzene_type} (f), separately.

Benzene is known to have delocalized electrons due to the $\pi$ bonds created by the six-membered ring.
Correspondingly, $\mu_R^{\tau}(\vec{r}_{\rm {surf}})$ is high throughout the benzene ring as shown in Fig.~\ref{fig:benzene_type} (d).
On the other hand, in the case of naphthalene shown in Fig.~\ref{fig:benzene_type} (e), the high $\mu_R^{\tau}(\vec{r}_{\rm surf})$ due to the double bond is seen at four corners at the upper and lower ends.
As easily expected,
the RCP $\mu_R^{\tau}(\vec{r}_{\rm {surf}})$ originating from the Lowest Unoccupied Molecular Orbital (LUMO) contribution shown in Fig.~\ref{fig:benzene_type} (f) is totally different from that originating from the HOMO contribution shown in Fig.~\ref{fig:benzene_type} (e).
Therefore, it is necessary to determine the appropriate energy window depending on whether the target substance is a donor or acceptor of electrons.
It is clear from the $\mu_R^{\tau}(\vec{r})$ distribution shown in Fig.~\ref{fig:benzene_type} (g) that pentacene has a bonding state between the outer central carbons similar to that of benzene rings, and four double bonding sites similar to those of naphthalene at the upper and lower ends, which is also consistent with the bond length argument.
When molecules interact with another inactive material surface such as an AFM probe with CO molecules attached, 
high $\mu_R^{\tau}(\vec{r}_{\rm {surf}})$ becomes a measure of the Pauli repulsive force originating from antibonding.
Figure \ref{fig:benzene_type} (g) is consistent with the topology image of non-contact AFM experiments using a CO probe \cite{Moll_2010,https://doi.org/10.1002/anie.201703509}.

\subsection{Benzene derivatives}
\label{sec:benzene_derivative}
\begin{figure}
    \centering
    \includegraphics[width=1.0\linewidth]{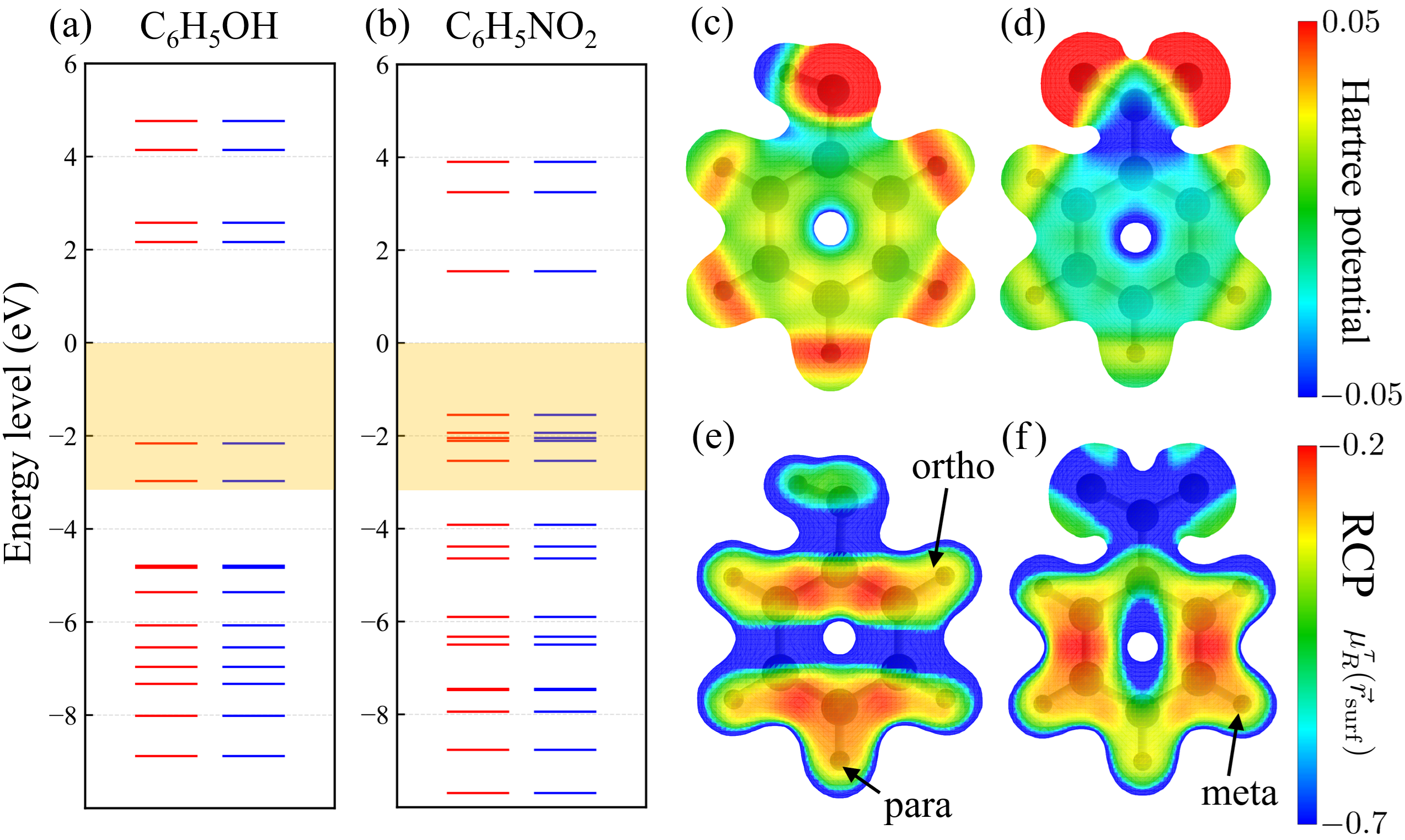}
    \caption{Energy levels of (a) phenol (C$_6$H$_5$OH) and (b) nitrobenzene (C$_6$H$_5$NO$_2$) molecules.
    The area colored in orange represents the energy window for RCP $\mu_R^{\tau}(\vec{r})$.
    The Hartree potential on the surface for (c) phenol and nitrobenzene (d) are drawn as color maps on the surface.
    The RCP $\mu_R^{\tau}(\vec{r})$ on the surface for (e) phenol and nitrobenzene (f) are drawn as color maps on the surface.}
    \label{fig:benzene_derivative}
\end{figure}

The energy levels, Hartree potentials at material surfaces, and the RCP $\mu_R^{\tau}(\vec{r}_{\rm {surf}})$ of phenol (C$_6$H$_5$OH)
 and nitrobenzene (C$_6$H$_5$NO$_2$) molecules
are shown in Fig.~\ref{fig:benzene_derivative}.
It is known that substituted benzenes exhibit distinct regioselectivities in electrophilic aromatic substitution reactions.
Phenol acts as an ortho/para-directing group due to the electron-donating effect of the hydroxyl substituent, whereas nitrobenzene is meta-directing as a result of the electron-withdrawing nature of the nitro group \cite{carey2007advanced}.
The electrostatic potential (Hartee potential), which is caused by intramolecular polarization, is a measure of Coulomb interaction with charged particles.
Although the Hartree potential is sometimes used for discussions of electron-donating ability for ionic materials,
it is not effective for regioselectivities in selectrophilic aromatic substitution reactions as shown in Figs.~\ref{fig:benzene_derivative} (c) and (d).
On the other hand, the local RCPs indicate the electron-donating ability at the surface.
The distribution of local RCPs clearly visualizes the ortho/para-directing of the phenol and the meta-directing of the nitrobenzene.
Here, we note that the energy windows were selected to include multiple orbitals, as it is known that the HOMO alone cannot account for the meta-directing behavior of nitrobenzene~\cite{doi:10.1021/jp027330l}.

\subsection{C(001) diamond surface}
\label{sec:C001}
\begin{figure}
    \centering
    \includegraphics[width=1\linewidth]{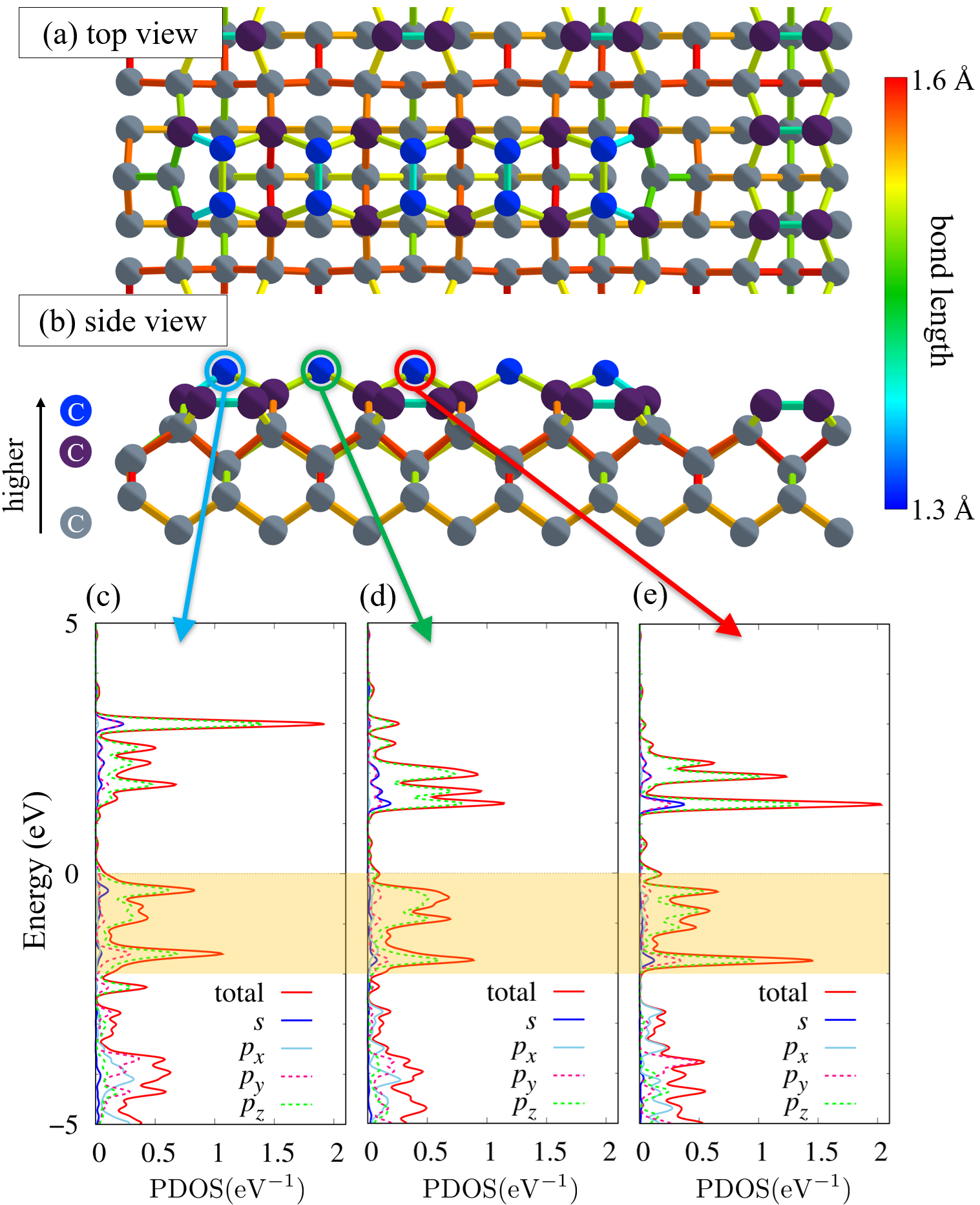}
    \caption{The top view (a) and side view (b) of a geometrical structure of the 5 carbon dimer ribbon on C(001) diamond surface.
    Colors of the bonds represent the bond length.
    (c)-(e) The partial DOSs of three C atoms of the dimer ribbon.
    The area colored in orange represents the energy window for RCP $\mu_R^{\tau}(\vec{r})$ in Fig.~\ref{fig:C001_2}.
    }
    \label{fig:C001}
\end{figure}

\begin{figure}
    \centering
    \includegraphics[width=1\linewidth]{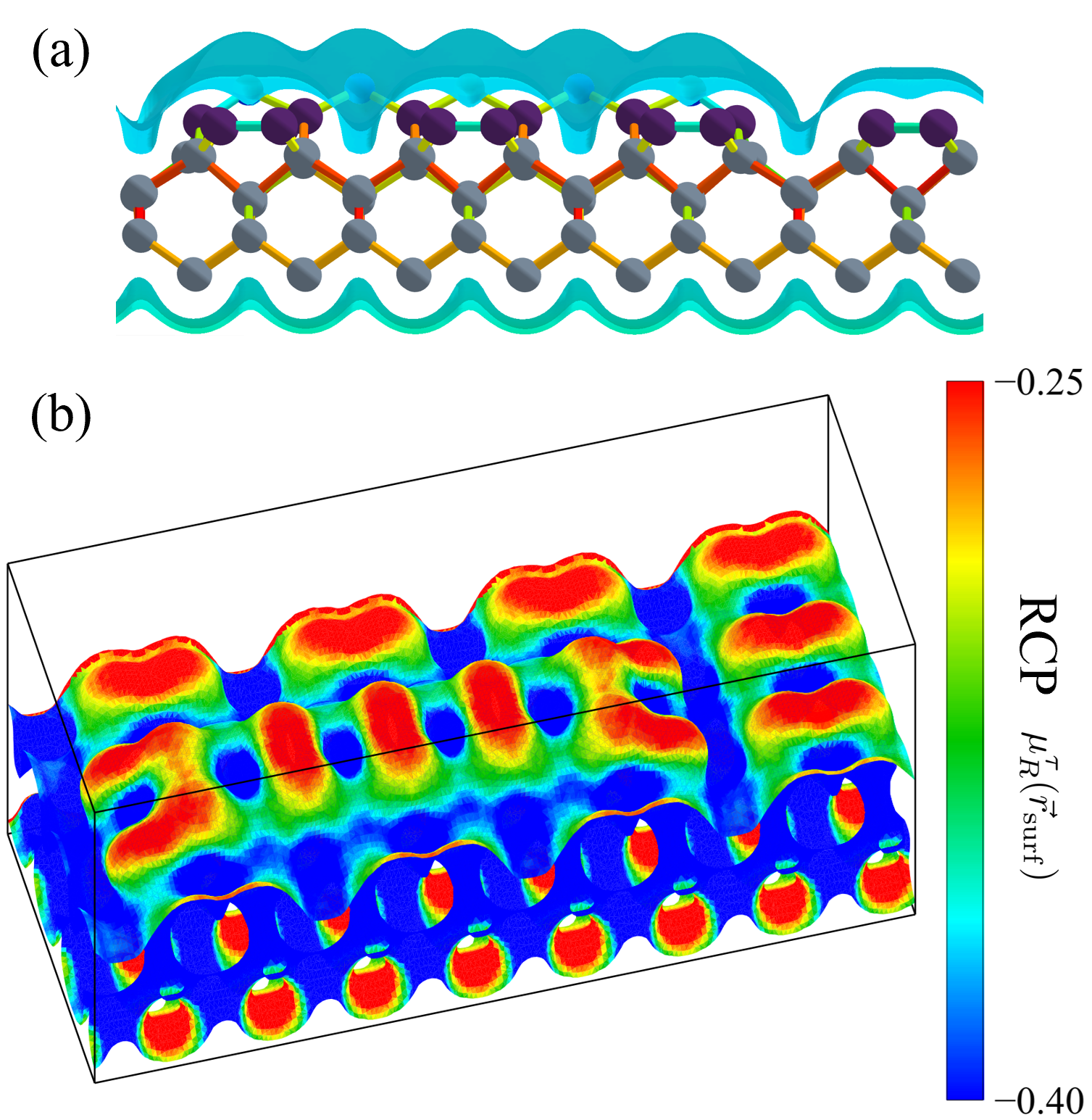}
    \caption{(a) The isosurface of electron density of the 5 carbon dimer ribbon on C(001) diamond surface in the energy range between -2\,eV and 0\,eV (approximated Fukui function). The isosurface value is 0.01 a.u.
    (b) The corresponding RCP $\mu_R^{\tau}(\vec{r})$ on the surface.
    The drawing area is cropped from the actual system.}
    \label{fig:C001_2}
\end{figure}

The dangling bonds and double bonds in molecules are usually analyzed using other properties such as the partial density of states, the electron charge density~\cite{10.1093/oso/9780198551683.001.0001}, the frontier orbitals (Fukui function)~\cite{10.1093/oso/9780195092769.001.0001}.
However, these properties are sometimes not sufficient to analyze the surface of the materials.
Figures \ref{fig:C001} (a) and (b) show a structure of the C dimer ribbon on the C(001) diamond surface, which was recently investigated in great detail by sophisticated AFM experiments~\cite{PhysRevResearch.7.023036,Zhang2025}.
The AFM images observed by the experiments revealed that the force between the AFM tip and the dimers of the ends of the C dimer ribbon is stronger than the force between the tip and the inner dimers~\cite{PhysRevResearch.7.023036}.
Figures \ref{fig:C001} (c)-(e) and Fig.~\ref{fig:C001_2} (a) show the partial density of states and electron charge density change $\Delta \rho(\vec{r})$, which is approximately proportional to the Fukui function $f(\vec{r}) \equiv \frac{\partial \rho(\vec{r})}{\partial N} \propto \Delta \rho(\vec{r})$.
There is little difference among C dimer atoms from the point of view of the partial density of states (PDOS) and the approximated Fukui function, so that they cannot give a clear picture of the electron-donating ability. 

However, the distribution of $\mu_R^{\tau}(\vec{r}_{\rm surf})$ as shown in Fig.~\ref{fig:C001_2} (b) reveals that each C atom of the end dimers forms a double bond with a C atom of the substrate below.  
This is consistent with the discussion of the bond length~\cite{PhysRevResearch.7.023036}.
The bond lengths of the corresponding double bond are approximately 1.38 - 1.40 \AA, while those of the other bonds in the dimer ribbon are approximately 1.50 \AA\ as shown in Figs.~\ref{fig:C001} (a) and (b).
The red region shown in Fig.~\ref{fig:C001_2} (b) corresponds to the high local RCP $\mu_R^{\tau}(\vec{r})$
originating from the C-C double bonds, which are highly electron-donating regions. Electrons are more likely to be supplied to outer regions in their electron-donating regions, which can form covalent bonding with an approaching tip of atomic force spectroscopy. As a result, higher-resolution AFM images are observed at the regions.
This result is consistent with the topology image of near-contact AFM experiments using an active Si probe~\cite{PhysRevResearch.7.023036,Zhang2025}.

\begin{figure*}
    \centering
    \includegraphics[width=0.8\linewidth]{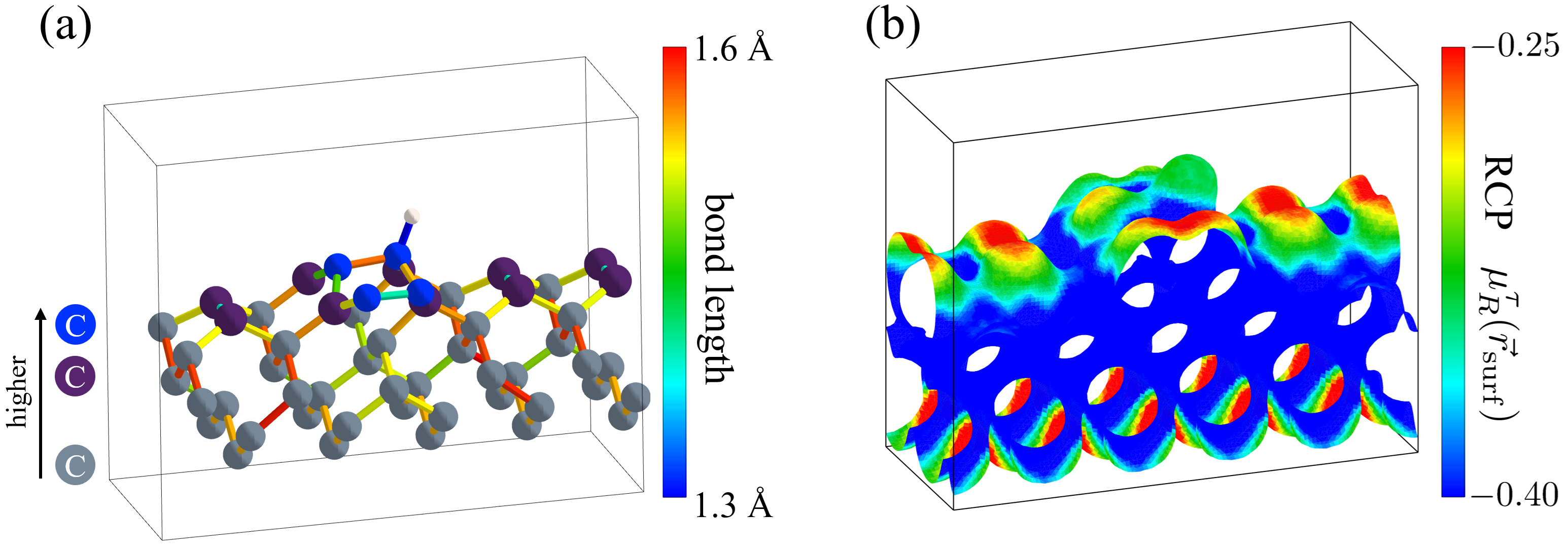}
    \caption{(a) The geometrical structure of the carbon dimer chain on C(001) diamond surface, where the one side of carbon dimer is hydrogenated.
    Colors of the bonds represent the bond length.
    (b) The corresponding RCP $\mu_R^{\tau}(\vec{r})$ on the surface.}
    \label{fig:C001_withH}
\end{figure*}

Finally, Figs.~\ref{fig:C001_withH} (a) and (b) show a structure and local RCP $\mu_R^{\tau}(\vec{r}_{\rm surf})$ of the periodic system of the hydrogenated C dimer chain on the C(001) surface.
The adsorption of a hydrogen atom on one of the carbon atoms in the dimer results in the formation of a dangling bond on the other carbon atom, which was originally involved in a double bond.
The double bonds and the dangling bond are clearly visualized in Fig.~\ref{fig:C001_withH} (b).

%
%

\section{Conclusions} \label{sec:Conclusions}
In this paper, we have proposed the local RCP analysis method using energy window scheme to estimate the ease of molecular adsorption on a material surface and the strength of the chemical bonding forces between an AFM tip and a material surface.

The local RCP at the electronic interface $\mu_R^{\tau}(\vec{r}_{\rm surf})$ is a measure of how much electron transfer from the surface. 
Two types of electron transfers have been discussed here: ``electron transfer from the material surface to the tip'' (ionic bonding picture) and ``electron transfer from the material surface or tip to the bonding region'' (covalent bonding picture).
Both cases of electron transfers are related to molecular adsorption and interactions between a tip and a surface.
In particular, we have focused on covalent bonding and discussed the relationship between the chemical bonding force and $\mu_R^{\tau}(\vec{r}_{\rm surf})$ using simple model calculations.
Furthermore, DFT calculations of several molecules and C(001) diamond surfaces have been performed to visualize spatial distribution of 
the RCPs $\mu_R^{\tau}(\vec{r}_{\rm surf})$.
As a result, we have successfully visualized dangling bonds and double bonds, which are regions with high electron-donating ability.
In other words, the RCPs $\mu_R^{\tau}(\vec{r}_{\rm surf})$ have visualized electron-donating ability at the surface where covalent bonds tend to be formed potentially.
These results indicate that the RCP $\mu_R^{\tau}(\vec{r}_{\rm surf})$ has potential to analyze high-resolution AFM images, which are observed by chemical bonding forces.
Furthermore, incorporating atomic height information and surface deformations that occur during AFM measurements remains an important subject for future work.

\section{Acknowledgement}
The computation in this work has been done using the facilities of
the Supercomputer Center, the Institute for Solid State Physics, the
University of Tokyo (ISSPkyodo-SC-2023-Ba-0073,2024-Ba-0053).
The work of M.S. was supported by Grants-in-Aid for Scientific
Research (22K12060).

\appendix
\section{Approximation for DFT}
\label{sec:Supp_Approx_DFT}
It is not easy to calculate the electronic physical quantities using a quantum field theoretical state vector.
Instead we approximate the physical quantities defined in the framework of the quantum field theory using density matrix obtained by two-component relativistic or non-relativistic DFT.

The electron field operator $\hat{\psi}(\vec{r}\,)$ can be expanded 
by two component expanding function $\psi_{i \vec{k}}(\vec{r}\,)$ and annihilation operator $\hat{e}_{i \vec{k}}$ as
\begin{align}
\hat{\psi}(\vec{r}) = \frac{1}{V_B} \int_{B} d^3\vec{k} \sum_{i=0}^\infty \psi_{i \vec{k}} (\vec{r}) \hat{e}_{i \vec{k}}, \label{eq:expand_psi}
\end{align}
where $i$ and $\vec{k}$ is the index of the energy state and a wave number vector, respectively. 
The integral $\int_{\rm B}$ means the integration over the first Brillouin zone of which volume is $V_{\rm B}$.
Using Eq.~\eqref{eq:expand_psi}, the electron number density operator,
$\hat{n} (\vec{r}) = \hat{\psi}^\dagger(\vec{r}) \hat{\psi}(\vec{r})$,
is expanded as
\begin{align}
\hat{n}(\vec{r}) = \frac{1}{V_B^2} \int_{B} d^3\vec{k} \int_{B} d^3\vec{k}' \sum_{i,i'=0}^\infty {\psi}_{i \vec{k}}^{\dagger} (\vec{r}){\psi}_{i' \vec{k}'}(\vec{r}) \hat{e}_{i \vec{k}}^{\dagger} \hat{e}_{i' \vec{k}'}.
\end{align}

The DFT approximation for the density matrix gives an approximated expectation value of the electron number density.
\begin{align}
\left\langle \hat{e}_{i \vec{k}}^{\dagger} \hat{e}_{i' \vec{k}'} \right\rangle &\approx
\left\langle \hat{e}_{i \vec{k}}^{\dagger} \hat{e}_{i' \vec{k}'} \right\rangle_{\rm DFT},\\
\left\langle \hat{e}_{i \vec{k}}^{\dagger} \hat{e}_{i' \vec{k}'} \right\rangle_{\rm DFT}& = \int_{-\infty}^{\infty} dE \delta_{ii'} \delta_{\vec{k} \vec{k}'} \delta(E- \epsilon_i) f(E, \mu, T),
\end{align}
where $\epsilon_i$ is the $i$-th eigenenergy.
The expectation value of the electron number density is given as
\begin{align}
n(\vec{r}) 
&\equiv \langle \hat{n}(\vec{r}) \rangle \approx \langle \hat{n}(\vec{r}) \rangle_{\rm DFT},
\end{align}
\small{
\begin{align}
\nonumber
 \left\langle \hat{n}(\vec{r}) \right\rangle_{\rm DFT} &= 
\frac{1}{V_B^2} \int_{B} d^3\vec{k} \int_{B} d^3\vec{k}' \sum_{i,i'=0}^\infty {\psi}_{i \vec{k}}^{\dagger} (\vec{r}){\psi}_{i' \vec{k}'}(\vec{r}) \notag \\
&\quad \times 
\left\langle \hat{e}_{i \vec{k}}^{\dagger} \hat{e}_{i' \vec{k}'} \right\rangle_{\rm DFT},\\
&= \int_{-\infty}^{\infty} dE n(\vec{r}, E) f(E, \mu, T),
\end{align}
}%
where
\begin{align}
n(\vec{r},E) &\equiv \frac{1}{V_B} \int_{B} d^3\vec{k} \sum_{i=0}^{\infty} {\psi}_{i \vec{k}}^{\dagger} (\vec{r}){\psi}_{i \vec{k}}(\vec{r}) \delta(E- \epsilon_i).
\end{align}
Similarly, the energy density is given as
\begin{align}
\varepsilon_\tau(\vec{r})  &\approx
 \int_{-\infty}^{\infty} dE \varepsilon_{\tau} (\vec{r}, E) f(E, \mu, T),\\
\varepsilon_{\tau} (\vec{r},E) &\equiv \frac{1}{V_B} \int_{B} d^3\vec{k} \sum_{i=0}^{\infty} \frac{1}{2} \sum_{l=1}^3 \tau^{Sll}_{e i \vec{k}} (\vec{r}) \delta(E- \epsilon_i),
\end{align}
where $\tau^{Slm}_{e i \vec{k}} (\vec{r})$ 
is expressed in a form similar to Eq.~\eqref{eq:stress} under PRQED approximation,
{\small 
\begin{align}
{\tau}^{Slm}_{ei\vec{k}} &\approx \frac{\hbar^2}{4m_e} \left[ {\psi}_{i\vec{k}}^{L \dagger} D_{el} D_{em}  {\psi}^L_{i\vec{k}} - \left(  D_{el} {\psi}^L_{i\vec{k}} \right)^\dagger  D_{em} {\psi}^L_{i\vec{k}}\right] + h.c.
\end{align}
}%

For the linear combination of the atomic orbital (LCAO) method,
the function $\psi^L_{i \vec{k}} (\vec{r})$ can be written by $\psi_{\sigma i}(\vec{k}, \vec{r})$,
which is expanded using expansion coefficients $ c_{\sigma i, p \alpha}(\vec{k}) $ and atomic orbitals $ \phi_{p \alpha}(\vec{r}) $
as follows.
\begin{align}
\psi^L_{i \vec{k}} (\vec{r}) = \begin{pmatrix}
    \psi_{\uparrow i}(\vec{k}, \vec{r}) \\
    \psi_{\downarrow i}(\vec{k}, \vec{r}) 
\end{pmatrix},
\end{align}
where
\begin{align}
\psi_{\sigma i}(\vec{k}, \vec{r}) = \frac{1}{N} \sum_{n}^{N} e^{i \vec{T}_n \cdot \vec{k}} \sum_{p, \alpha} c_{\sigma i, p \alpha}(\vec{k}) \, \phi_{p \alpha}(\vec{r} - \vec{R}_p - \vec{T}_n).
\end{align}
Here, $ \vec{T}_n $ is a lattice vector, $N$ is the number of a cell, $ p (q)$ is a site index corresponding to a position $ \vec{R}_p (\vec{R}_q)$, $ \sigma $ (either $ \uparrow $ or $ \downarrow $) denotes the spin index, and $ \alpha$ and $\beta$ are the orbital indices.
Then, Eq.~\eqref{eq:stress} ignoring the vector potential is expanded as
\small{
\begin{align}
{\tau}^{Slm}_{ei\vec{k}} (\vec{r})&\approx \frac{\hbar^2}{4m_e N} \sum_n^N \sum_\sigma \rho_{\sigma \sigma, p \alpha q \beta}^{(\vec{T}_n)} (\vec{k}) \sum_j^2 \omega^{(j) lm(\vec{T}_n)}_{p \alpha q \beta} (\vec{r}) + h.c., \label{eq:stress_LCAO}
\end{align}
}%
where
\begin{align}
\rho_{\sigma \sigma', p \alpha q \beta}^{(\vec{T}_n)} (\vec{k}) &\equiv  e^{i \vec{T}_n \cdot \vec{k}} c_{\sigma i, p \alpha}^{*} (\vec{k}) c_{\sigma' i, q \beta}(\vec{k}), \\
\omega^{(1) lm(\vec{T}_n)}_{p \alpha q \beta} (\vec{r}) &\equiv \phi^*_{p \alpha} \left(\vec{r}-\vec{R}_p \right) \partial_l \partial_m \phi_{q \beta} \left(\vec{r} - \vec{R}_q - \vec{T}_n \right), \\
\omega^{(2) lm(\vec{T}_n)}_{p \alpha q \beta} (\vec{r}) &\equiv -\partial_l \phi^*_{p \alpha} \left(\vec{r}-\vec{R}_p \right) \partial_m \phi_{q \beta} \left(\vec{r} - \vec{R}_q - \vec{T}_n \right).
\end{align}

For the plane wave (PW) method, 
$\psi_{\sigma i}(\vec{k}, \vec{r})$ is expanded by the coefficient $C_{\sigma i, \vec{G}} (\vec{k}) $ and the plane wave $ e^{i \vec{k} \cdot \vec{r}} $ as
\begin{align}
    \psi_{\sigma i}(\vec{k}, \vec{r}) = \frac{1}{\sqrt{\Omega}} \sum_{\vec{G}} C_{\sigma i, \vec{G}} (\vec{k}) e^{i (\vec{k} + \vec{G}) \cdot \vec{r}},
\end{align}
where $\vec{G}$ is a reciprocal lattice vector, and $\Omega$ is a normalization factor.
Then, Eq.~\eqref{eq:stress} ignoring the vector potential is expanded as
\begin{align}
{\tau}^{Slm}_{ei\vec{k}} (\vec{r}) &\approx \frac{\hbar^2}{4m_e \sqrt{\Omega}} \sum_{\vec{G},\vec{G}'} \sum_\sigma \rho_{\sigma \sigma}^{\vec{G}, \vec{G'}} (\vec{k}) \omega^{lm}_{\vec{k},\vec{G}, \vec{G}'} (\vec{r}) + h.c.,
\end{align}
where
\begin{align}
\rho_{\sigma \sigma'}^{\vec{G}, \vec{G}'} (\vec{k}) &\equiv  C_{\sigma i, \vec{G}}^{*} (\vec{k}) C_{\sigma' i, \vec{G}'}(\vec{k}),\\
\omega^{(1)lm}_{\vec{k},\vec{G}, \vec{G}'} (\vec{r}) &\equiv -(k^l + G'^l)(k^m+G'^m) e^{i (\vec{G}'-\vec{G}) \cdot \vec{r}}, \\
\omega^{(2)lm}_{\vec{k},\vec{G}, \vec{G}'} (\vec{r}) &\equiv -(k^l + G^l)(k^m+G'^m) e^{i (\vec{G}'-\vec{G}) \cdot \vec{r}},\\
\omega^{lm}_{\vec{k},\vec{G}, \vec{G}'} (\vec{r}) &\equiv \omega^{(1) lm}_{\vec{k},\vec{G}, \vec{G}'} (\vec{r}) + \omega^{(2) lm}_{\vec{k},\vec{G}, \vec{G}'} (\vec{r}), \\
&= -(2k^l + G^l + G'^l)(k^m+G'^m) e^{i (\vec{G}'-\vec{G}) \cdot \vec{r}}.
\end{align}

In this study, we used Eq.~\eqref{eq:stress_LCAO} to calculate the RCP based on the LCAO method by interpreting the atomic orbitals as local pseudo-atomic basis functions.


\bibliography{ref}

\begin{thebibliography}{49}%
\makeatletter
\providecommand \@ifxundefined [1]{%
 \@ifx{#1\undefined}
}%
\providecommand \@ifnum [1]{%
 \ifnum #1\expandafter \@firstoftwo
 \else \expandafter \@secondoftwo
 \fi
}%
\providecommand \@ifx [1]{%
 \ifx #1\expandafter \@firstoftwo
 \else \expandafter \@secondoftwo
 \fi
}%
\providecommand \natexlab [1]{#1}%
\providecommand \enquote  [1]{``#1''}%
\providecommand \bibnamefont  [1]{#1}%
\providecommand \bibfnamefont [1]{#1}%
\providecommand \citenamefont [1]{#1}%
\providecommand \href@noop [0]{\@secondoftwo}%
\providecommand \href [0]{\begingroup \@sanitize@url \@href}%
\providecommand \@href[1]{\@@startlink{#1}\@@href}%
\providecommand \@@href[1]{\endgroup#1\@@endlink}%
\providecommand \@sanitize@url [0]{\catcode `\\12\catcode `\$12\catcode `\&12\catcode `\#12\catcode `\^12\catcode `\_12\catcode `\%12\relax}%
\providecommand \@@startlink[1]{}%
\providecommand \@@endlink[0]{}%
\providecommand \url  [0]{\begingroup\@sanitize@url \@url }%
\providecommand \@url [1]{\endgroup\@href {#1}{\urlprefix }}%
\providecommand \urlprefix  [0]{URL }%
\providecommand \Eprint [0]{\href }%
\providecommand \doibase [0]{https://doi.org/}%
\providecommand \selectlanguage [0]{\@gobble}%
\providecommand \bibinfo  [0]{\@secondoftwo}%
\providecommand \bibfield  [0]{\@secondoftwo}%
\providecommand \translation [1]{[#1]}%
\providecommand \BibitemOpen [0]{}%
\providecommand \bibitemStop [0]{}%
\providecommand \bibitemNoStop [0]{.\EOS\space}%
\providecommand \EOS [0]{\spacefactor3000\relax}%
\providecommand \BibitemShut  [1]{\csname bibitem#1\endcsname}%
\let\auto@bib@innerbib\@empty
\bibitem [{\citenamefont {Parr}\ and\ \citenamefont {Yang}(1989)}]{parr1989}%
  \BibitemOpen
  \bibfield  {author} {\bibinfo {author} {\bibfnamefont {R.~G.}\ \bibnamefont {Parr}}\ and\ \bibinfo {author} {\bibfnamefont {W.}~\bibnamefont {Yang}},\ }\href@noop {} {\emph {\bibinfo {title} {Density-Functional Theory of Atoms and Molecules}}}\ (\bibinfo  {publisher} {Oxford University Press},\ \bibinfo {address} {New York, Oxford},\ \bibinfo {year} {1989})\BibitemShut {NoStop}%
\bibitem [{\citenamefont {Tachibana}\ and\ \citenamefont {Parr}(1992)}]{Tachibana1992}%
  \BibitemOpen
  \bibfield  {author} {\bibinfo {author} {\bibfnamefont {A.}~\bibnamefont {Tachibana}}\ and\ \bibinfo {author} {\bibfnamefont {R.~G.}\ \bibnamefont {Parr}},\ }\bibfield  {title} {\bibinfo {title} {On the redistribution of electrons for chemical reaction systems},\ }\href {https://doi.org/https://doi.org/10.1002/qua.560410402} {\bibfield  {journal} {\bibinfo  {journal} {International Journal of Quantum Chemistry}\ }\textbf {\bibinfo {volume} {41}},\ \bibinfo {pages} {527} (\bibinfo {year} {1992})}\BibitemShut {NoStop}%
\bibitem [{\citenamefont {Herring}\ and\ \citenamefont {Nichols}(1949)}]{RevModPhys.21.185}%
  \BibitemOpen
  \bibfield  {author} {\bibinfo {author} {\bibfnamefont {C.}~\bibnamefont {Herring}}\ and\ \bibinfo {author} {\bibfnamefont {M.~H.}\ \bibnamefont {Nichols}},\ }\bibfield  {title} {\bibinfo {title} {Thermionic emission},\ }\href {https://doi.org/10.1103/RevModPhys.21.185} {\bibfield  {journal} {\bibinfo  {journal} {Rev. Mod. Phys.}\ }\textbf {\bibinfo {volume} {21}},\ \bibinfo {pages} {185} (\bibinfo {year} {1949})}\BibitemShut {NoStop}%
\bibitem [{\citenamefont {Tachibana}\ \emph {et~al.}(1999)\citenamefont {Tachibana}, \citenamefont {Nakamura}, \citenamefont {Sakata},\ and\ \citenamefont {Morisaki}}]{tachibana1999}%
  \BibitemOpen
  \bibfield  {author} {\bibinfo {author} {\bibfnamefont {A.}~\bibnamefont {Tachibana}}, \bibinfo {author} {\bibfnamefont {K.}~\bibnamefont {Nakamura}}, \bibinfo {author} {\bibfnamefont {K.}~\bibnamefont {Sakata}},\ and\ \bibinfo {author} {\bibfnamefont {T.}~\bibnamefont {Morisaki}},\ }\bibfield  {title} {\bibinfo {title} {Application of the regional density functional theory: The chemical potential inequality in the heh+ system},\ }\href {https://doi.org/https://doi.org/10.1002/(SICI)1097-461X(1999)74:6<669::AID-QUA8>3.0.CO;2-O} {\bibfield  {journal} {\bibinfo  {journal} {International Journal of Quantum Chemistry}\ }\textbf {\bibinfo {volume} {74}},\ \bibinfo {pages} {669} (\bibinfo {year} {1999})}\BibitemShut {NoStop}%
\bibitem [{\citenamefont {Szarek}\ and\ \citenamefont {Tachibana}(2007)}]{Szarek2007}%
  \BibitemOpen
  \bibfield  {author} {\bibinfo {author} {\bibfnamefont {P.}~\bibnamefont {Szarek}}\ and\ \bibinfo {author} {\bibfnamefont {A.}~\bibnamefont {Tachibana}},\ }\bibfield  {title} {\bibinfo {title} {The field theoretical study of chemical interaction in terms of the rigged qed: new reactivity indices},\ }\href@noop {} {\bibfield  {journal} {\bibinfo  {journal} {J. Mol. Model.}\ }\textbf {\bibinfo {volume} {13}},\ \bibinfo {pages} {651} (\bibinfo {year} {2007})}\BibitemShut {NoStop}%
\bibitem [{\citenamefont {Szarek}\ \emph {et~al.}(2008)\citenamefont {Szarek}, \citenamefont {Sueda},\ and\ \citenamefont {Tachibana}}]{10.1063/1.2973634}%
  \BibitemOpen
  \bibfield  {author} {\bibinfo {author} {\bibfnamefont {P.}~\bibnamefont {Szarek}}, \bibinfo {author} {\bibfnamefont {Y.}~\bibnamefont {Sueda}},\ and\ \bibinfo {author} {\bibfnamefont {A.}~\bibnamefont {Tachibana}},\ }\bibfield  {title} {\bibinfo {title} {{Electronic stress tensor description of chemical bonds using nonclassical bond order concept}},\ }\href {https://doi.org/10.1063/1.2973634} {\bibfield  {journal} {\bibinfo  {journal} {The Journal of Chemical Physics}\ }\textbf {\bibinfo {volume} {129}},\ \bibinfo {pages} {094102} (\bibinfo {year} {2008})}\BibitemShut {NoStop}%
\bibitem [{\citenamefont {Szarek}\ \emph {et~al.}(2009)\citenamefont {Szarek}, \citenamefont {Urakami}, \citenamefont {Zhou}, \citenamefont {Cheng},\ and\ \citenamefont {Tachibana}}]{10.1063/1.3072369}%
  \BibitemOpen
  \bibfield  {author} {\bibinfo {author} {\bibfnamefont {P.}~\bibnamefont {Szarek}}, \bibinfo {author} {\bibfnamefont {K.}~\bibnamefont {Urakami}}, \bibinfo {author} {\bibfnamefont {C.}~\bibnamefont {Zhou}}, \bibinfo {author} {\bibfnamefont {H.}~\bibnamefont {Cheng}},\ and\ \bibinfo {author} {\bibfnamefont {A.}~\bibnamefont {Tachibana}},\ }\bibfield  {title} {\bibinfo {title} {{On reversible bonding of hydrogen molecules on platinum clusters}},\ }\href {https://doi.org/10.1063/1.3072369} {\bibfield  {journal} {\bibinfo  {journal} {The Journal of Chemical Physics}\ }\textbf {\bibinfo {volume} {130}},\ \bibinfo {pages} {084111} (\bibinfo {year} {2009})}\BibitemShut {NoStop}%
\bibitem [{\citenamefont {Senami}\ \emph {et~al.}(2011)\citenamefont {Senami}, \citenamefont {Ikeda}, \citenamefont {Fukushima},\ and\ \citenamefont {Tachibana}}]{10.1063/1.3651182}%
  \BibitemOpen
  \bibfield  {author} {\bibinfo {author} {\bibfnamefont {M.}~\bibnamefont {Senami}}, \bibinfo {author} {\bibfnamefont {Y.}~\bibnamefont {Ikeda}}, \bibinfo {author} {\bibfnamefont {A.}~\bibnamefont {Fukushima}},\ and\ \bibinfo {author} {\bibfnamefont {A.}~\bibnamefont {Tachibana}},\ }\bibfield  {title} {\bibinfo {title} {Theoretical study of adsorption of lithium atom on carbon nanotube},\ }\href {https://doi.org/10.1063/1.3651182} {\bibfield  {journal} {\bibinfo  {journal} {AIP Advances}\ }\textbf {\bibinfo {volume} {1}},\ \bibinfo {pages} {042106} (\bibinfo {year} {2011})}\BibitemShut {NoStop}%
\bibitem [{\citenamefont {Onoda}\ \emph {et~al.}(2020)\citenamefont {Onoda}, \citenamefont {Miyazaki},\ and\ \citenamefont {Sugimoto}}]{doi:10.1021/acs.nanolett.9b05280}%
  \BibitemOpen
  \bibfield  {author} {\bibinfo {author} {\bibfnamefont {J.}~\bibnamefont {Onoda}}, \bibinfo {author} {\bibfnamefont {H.}~\bibnamefont {Miyazaki}},\ and\ \bibinfo {author} {\bibfnamefont {Y.}~\bibnamefont {Sugimoto}},\ }\bibfield  {title} {\bibinfo {title} {Chemical identification of the foremost tip atom in atomic force microscopy},\ }\href {https://doi.org/10.1021/acs.nanolett.9b05280} {\bibfield  {journal} {\bibinfo  {journal} {Nano Letters}\ }\textbf {\bibinfo {volume} {20}},\ \bibinfo {pages} {2000} (\bibinfo {year} {2020})},\ \bibinfo {note} {pMID: 32031816}\BibitemShut {NoStop}%
\bibitem [{\citenamefont {Tachibana}(2001)}]{Tachibana2001}%
  \BibitemOpen
  \bibfield  {author} {\bibinfo {author} {\bibfnamefont {A.}~\bibnamefont {Tachibana}},\ }\bibfield  {title} {\bibinfo {title} {Electronic energy density in chemical reaction systems},\ }\href {https://doi.org/10.1063/1.1384012} {\bibfield  {journal} {\bibinfo  {journal} {The Journal of Chemical Physics}\ }\textbf {\bibinfo {volume} {115}},\ \bibinfo {pages} {3497} (\bibinfo {year} {2001})}\BibitemShut {NoStop}%
\bibitem [{\citenamefont {Sanderson}(1951)}]{doi:10.1126/science.114.2973.670}%
  \BibitemOpen
  \bibfield  {author} {\bibinfo {author} {\bibfnamefont {R.~T.}\ \bibnamefont {Sanderson}},\ }\bibfield  {title} {\bibinfo {title} {An interpretation of bond lengths and a classification of bonds},\ }\href {https://doi.org/10.1126/science.114.2973.670} {\bibfield  {journal} {\bibinfo  {journal} {Science}\ }\textbf {\bibinfo {volume} {114}},\ \bibinfo {pages} {670} (\bibinfo {year} {1951})}\BibitemShut {NoStop}%
\bibitem [{\citenamefont {Nozaki}\ \emph {et~al.}(2016)\citenamefont {Nozaki}, \citenamefont {Ichikawa},\ and\ \citenamefont {Tachibana}}]{https://doi.org/10.1002/qua.25073}%
  \BibitemOpen
  \bibfield  {author} {\bibinfo {author} {\bibfnamefont {H.}~\bibnamefont {Nozaki}}, \bibinfo {author} {\bibfnamefont {K.}~\bibnamefont {Ichikawa}},\ and\ \bibinfo {author} {\bibfnamefont {A.}~\bibnamefont {Tachibana}},\ }\bibfield  {title} {\bibinfo {title} {Theoretical study of atoms by the electronic kinetic energy density and stress tensor density},\ }\href {https://doi.org/https://doi.org/10.1002/qua.25073} {\bibfield  {journal} {\bibinfo  {journal} {International Journal of Quantum Chemistry}\ }\textbf {\bibinfo {volume} {116}},\ \bibinfo {pages} {504} (\bibinfo {year} {2016})}\BibitemShut {NoStop}%
\bibitem [{\citenamefont {Weinberg}(1972)}]{weinberg1972gravitation}%
  \BibitemOpen
  \bibfield  {author} {\bibinfo {author} {\bibfnamefont {S.}~\bibnamefont {Weinberg}},\ }\href@noop {} {\emph {\bibinfo {title} {Gravitation and Cosmology: Principles and Applications of the General Theory of Relativity}}}\ (\bibinfo  {publisher} {Wiley},\ \bibinfo {address} {New York},\ \bibinfo {year} {1972})\BibitemShut {NoStop}%
\bibitem [{\citenamefont {Tachibana}(2012)}]{Tachibana2012}%
  \BibitemOpen
  \bibfield  {author} {\bibinfo {author} {\bibfnamefont {A.}~\bibnamefont {Tachibana}},\ }\bibfield  {title} {\bibinfo {title} {General relativistic symmetry of electron spin torque},\ }\href {https://doi.org/10.1007/s10910-011-9943-z} {\bibfield  {journal} {\bibinfo  {journal} {Journal of Mathematical Chemistry}\ }\textbf {\bibinfo {volume} {50}},\ \bibinfo {pages} {669} (\bibinfo {year} {2012})}\BibitemShut {NoStop}%
\bibitem [{\citenamefont {Tachibana}(2014)}]{Tachibana2014}%
  \BibitemOpen
  \bibfield  {author} {\bibinfo {author} {\bibfnamefont {A.}~\bibnamefont {Tachibana}},\ }\bibfield  {title} {\bibinfo {title} {Stress tensor of electron as energy density with spin vorticity},\ }\href {https://doi.org/10.2477/jccj.2013-0012} {\bibfield  {journal} {\bibinfo  {journal} {Journal of Computer Chemistry, Japan}\ }\textbf {\bibinfo {volume} {13}},\ \bibinfo {pages} {18} (\bibinfo {year} {2014})}\BibitemShut {NoStop}%
\bibitem [{\citenamefont {Fukuda}\ \emph {et~al.}(2016{\natexlab{a}})\citenamefont {Fukuda}, \citenamefont {Soga}, \citenamefont {Senami},\ and\ \citenamefont {Tachibana}}]{Fukuda2016_LPQ}%
  \BibitemOpen
  \bibfield  {author} {\bibinfo {author} {\bibfnamefont {M.}~\bibnamefont {Fukuda}}, \bibinfo {author} {\bibfnamefont {K.}~\bibnamefont {Soga}}, \bibinfo {author} {\bibfnamefont {M.}~\bibnamefont {Senami}},\ and\ \bibinfo {author} {\bibfnamefont {A.}~\bibnamefont {Tachibana}},\ }\bibfield  {title} {\bibinfo {title} {Local physical quantities for spin based on the relativistic quantum field theory in molecular systems},\ }\href {https://doi.org/https://doi.org/10.1002/qua.25102} {\bibfield  {journal} {\bibinfo  {journal} {International Journal of Quantum Chemistry}\ }\textbf {\bibinfo {volume} {116}},\ \bibinfo {pages} {920} (\bibinfo {year} {2016}{\natexlab{a}})}\BibitemShut {NoStop}%
\bibitem [{\citenamefont {Fukuda}\ \emph {et~al.}(2016{\natexlab{b}})\citenamefont {Fukuda}, \citenamefont {Ichikawa}, \citenamefont {Senami},\ and\ \citenamefont {Tachibana}}]{fukuda2016_spinvorticity}%
  \BibitemOpen
  \bibfield  {author} {\bibinfo {author} {\bibfnamefont {M.}~\bibnamefont {Fukuda}}, \bibinfo {author} {\bibfnamefont {K.}~\bibnamefont {Ichikawa}}, \bibinfo {author} {\bibfnamefont {M.}~\bibnamefont {Senami}},\ and\ \bibinfo {author} {\bibfnamefont {A.}~\bibnamefont {Tachibana}},\ }\bibfield  {title} {\bibinfo {title} {Dynamical picture of spin hall effect based on quantum spin vorticity theory},\ }\href {https://doi.org/10.1063/1.4942087} {\bibfield  {journal} {\bibinfo  {journal} {AIP Advances}\ }\textbf {\bibinfo {volume} {6}},\ \bibinfo {pages} {025108} (\bibinfo {year} {2016}{\natexlab{b}})}\BibitemShut {NoStop}%
\bibitem [{\citenamefont {Tachibana}(2017)}]{Tachibana2017}%
  \BibitemOpen
  \bibfield  {author} {\bibinfo {author} {\bibfnamefont {A.}~\bibnamefont {Tachibana}},\ }\href {https://doi.org/10.1007/978-981-10-3132-8} {\emph {\bibinfo {title} {New Aspects of Quantum Electrodynamics}}}\ (\bibinfo  {publisher} {Springer Singapore},\ \bibinfo {year} {2017})\BibitemShut {NoStop}%
\bibitem [{\citenamefont {Tachibana}(2004)}]{Tachibana2004}%
  \BibitemOpen
  \bibfield  {author} {\bibinfo {author} {\bibfnamefont {A.}~\bibnamefont {Tachibana}},\ }\bibfield  {title} {\bibinfo {title} {Spindle structure of the stress tensor of chemical bond},\ }\href {https://doi.org/https://doi.org/10.1002/qua.20258} {\bibfield  {journal} {\bibinfo  {journal} {International Journal of Quantum Chemistry}\ }\textbf {\bibinfo {volume} {100}},\ \bibinfo {pages} {981} (\bibinfo {year} {2004})}\BibitemShut {NoStop}%
\bibitem [{\citenamefont {Zhang}\ \emph {et~al.}(2025{\natexlab{a}})\citenamefont {Zhang}, \citenamefont {Yasui}, \citenamefont {Fukuda}, \citenamefont {Ozaki}, \citenamefont {Ogura}, \citenamefont {Makino}, \citenamefont {Takeuchi},\ and\ \citenamefont {Sugimoto}}]{Zhang2025}%
  \BibitemOpen
  \bibfield  {author} {\bibinfo {author} {\bibfnamefont {R.}~\bibnamefont {Zhang}}, \bibinfo {author} {\bibfnamefont {Y.}~\bibnamefont {Yasui}}, \bibinfo {author} {\bibfnamefont {M.}~\bibnamefont {Fukuda}}, \bibinfo {author} {\bibfnamefont {T.}~\bibnamefont {Ozaki}}, \bibinfo {author} {\bibfnamefont {M.}~\bibnamefont {Ogura}}, \bibinfo {author} {\bibfnamefont {T.}~\bibnamefont {Makino}}, \bibinfo {author} {\bibfnamefont {D.}~\bibnamefont {Takeuchi}},\ and\ \bibinfo {author} {\bibfnamefont {Y.}~\bibnamefont {Sugimoto}},\ }\bibfield  {title} {\bibinfo {title} {Atomic observation on diamond (001) surfaces with near-contact atomic force microscopy},\ }\href {https://doi.org/10.1021/acs.nanolett.4c05395} {\bibfield  {journal} {\bibinfo  {journal} {Nano Letters}\ }\textbf {\bibinfo {volume} {25}},\ \bibinfo {pages} {1101} (\bibinfo {year} {2025}{\natexlab{a}})}\BibitemShut {NoStop}%
\bibitem [{\citenamefont {Zhang}\ \emph {et~al.}(2025{\natexlab{b}})\citenamefont {Zhang}, \citenamefont {Yasui}, \citenamefont {Fukuda}, \citenamefont {Ogura}, \citenamefont {Makino}, \citenamefont {Takeuchi}, \citenamefont {Ozaki},\ and\ \citenamefont {Sugimoto}}]{PhysRevResearch.7.023036}%
  \BibitemOpen
  \bibfield  {author} {\bibinfo {author} {\bibfnamefont {R.}~\bibnamefont {Zhang}}, \bibinfo {author} {\bibfnamefont {Y.}~\bibnamefont {Yasui}}, \bibinfo {author} {\bibfnamefont {M.}~\bibnamefont {Fukuda}}, \bibinfo {author} {\bibfnamefont {M.}~\bibnamefont {Ogura}}, \bibinfo {author} {\bibfnamefont {T.}~\bibnamefont {Makino}}, \bibinfo {author} {\bibfnamefont {D.}~\bibnamefont {Takeuchi}}, \bibinfo {author} {\bibfnamefont {T.}~\bibnamefont {Ozaki}},\ and\ \bibinfo {author} {\bibfnamefont {Y.}~\bibnamefont {Sugimoto}},\ }\bibfield  {title} {\bibinfo {title} {Dimer ribbon structures on diamond (001) surfaces revealed with atomic force microscopy},\ }\href {https://doi.org/10.1103/PhysRevResearch.7.023036} {\bibfield  {journal} {\bibinfo  {journal} {Phys. Rev. Res.}\ }\textbf {\bibinfo {volume} {7}},\ \bibinfo {pages} {023036} (\bibinfo {year} {2025}{\natexlab{b}})}\BibitemShut {NoStop}%
\bibitem [{\citenamefont {Moll}\ \emph {et~al.}(2010)\citenamefont {Moll}, \citenamefont {Gross}, \citenamefont {Mohn}, \citenamefont {Curioni},\ and\ \citenamefont {Meyer}}]{Moll_2010}%
  \BibitemOpen
  \bibfield  {author} {\bibinfo {author} {\bibfnamefont {N.}~\bibnamefont {Moll}}, \bibinfo {author} {\bibfnamefont {L.}~\bibnamefont {Gross}}, \bibinfo {author} {\bibfnamefont {F.}~\bibnamefont {Mohn}}, \bibinfo {author} {\bibfnamefont {A.}~\bibnamefont {Curioni}},\ and\ \bibinfo {author} {\bibfnamefont {G.}~\bibnamefont {Meyer}},\ }\bibfield  {title} {\bibinfo {title} {The mechanisms underlying the enhanced resolution of atomic force microscopy with functionalized tips},\ }\href {https://doi.org/10.1088/1367-2630/12/12/125020} {\bibfield  {journal} {\bibinfo  {journal} {New Journal of Physics}\ }\textbf {\bibinfo {volume} {12}},\ \bibinfo {pages} {125020} (\bibinfo {year} {2010})}\BibitemShut {NoStop}%
\bibitem [{\citenamefont {Tachibana}(2005)}]{Tachibana2005}%
  \BibitemOpen
  \bibfield  {author} {\bibinfo {author} {\bibfnamefont {A.}~\bibnamefont {Tachibana}},\ }\bibfield  {title} {\bibinfo {title} {A new visualization scheme of chemical energy density and bonds in molecules},\ }\href {https://doi.org/10.1007/s00894-005-0260-y} {\bibfield  {journal} {\bibinfo  {journal} {Journal of Molecular Modeling}\ }\textbf {\bibinfo {volume} {11}},\ \bibinfo {pages} {301} (\bibinfo {year} {2005})}\BibitemShut {NoStop}%
\bibitem [{\citenamefont {Hoffmann}(1963)}]{10.1063/1.1734456}%
  \BibitemOpen
  \bibfield  {author} {\bibinfo {author} {\bibfnamefont {R.}~\bibnamefont {Hoffmann}},\ }\bibfield  {title} {\bibinfo {title} {{An Extended Huckel Theory. I. Hydrocarbons}},\ }\href {https://doi.org/10.1063/1.1734456} {\bibfield  {journal} {\bibinfo  {journal} {The Journal of Chemical Physics}\ }\textbf {\bibinfo {volume} {39}},\ \bibinfo {pages} {1397} (\bibinfo {year} {1963})}\BibitemShut {NoStop}%
\bibitem [{\citenamefont {Kohn}\ and\ \citenamefont {Sham}(1965)}]{PhysRev.140.A1133}%
  \BibitemOpen
  \bibfield  {author} {\bibinfo {author} {\bibfnamefont {W.}~\bibnamefont {Kohn}}\ and\ \bibinfo {author} {\bibfnamefont {L.~J.}\ \bibnamefont {Sham}},\ }\bibfield  {title} {\bibinfo {title} {Self-consistent equations including exchange and correlation effects},\ }\href {https://doi.org/10.1103/PhysRev.140.A1133} {\bibfield  {journal} {\bibinfo  {journal} {Phys. Rev.}\ }\textbf {\bibinfo {volume} {140}},\ \bibinfo {pages} {A1133} (\bibinfo {year} {1965})}\BibitemShut {NoStop}%
\bibitem [{\citenamefont {Perdew}\ \emph {et~al.}(1996)\citenamefont {Perdew}, \citenamefont {Burke},\ and\ \citenamefont {Ernzerhof}}]{PhysRevLett.77.3865}%
  \BibitemOpen
  \bibfield  {author} {\bibinfo {author} {\bibfnamefont {J.~P.}\ \bibnamefont {Perdew}}, \bibinfo {author} {\bibfnamefont {K.}~\bibnamefont {Burke}},\ and\ \bibinfo {author} {\bibfnamefont {M.}~\bibnamefont {Ernzerhof}},\ }\bibfield  {title} {\bibinfo {title} {Generalized gradient approximation made simple},\ }\href {https://doi.org/10.1103/PhysRevLett.77.3865} {\bibfield  {journal} {\bibinfo  {journal} {Phys. Rev. Lett.}\ }\textbf {\bibinfo {volume} {77}},\ \bibinfo {pages} {3865} (\bibinfo {year} {1996})}\BibitemShut {NoStop}%
\bibitem [{\citenamefont {Ozaki}\ \emph {et~al.}()\citenamefont {Ozaki} \emph {et~al.}}]{OpenMX}%
  \BibitemOpen
  \bibfield  {author} {\bibinfo {author} {\bibfnamefont {T.}~\bibnamefont {Ozaki}} \emph {et~al.},\ }\href@noop {} {\bibinfo {title} {\textsc{OpenMX} package}},\ \bibinfo {note} {\url{http://www.openmx-square.org/}}\BibitemShut {NoStop}%
\bibitem [{\citenamefont {Morrison}\ \emph {et~al.}(1993)\citenamefont {Morrison}, \citenamefont {Bylander},\ and\ \citenamefont {Kleinman}}]{PhysRevB.47.6728}%
  \BibitemOpen
  \bibfield  {author} {\bibinfo {author} {\bibfnamefont {I.}~\bibnamefont {Morrison}}, \bibinfo {author} {\bibfnamefont {D.~M.}\ \bibnamefont {Bylander}},\ and\ \bibinfo {author} {\bibfnamefont {L.}~\bibnamefont {Kleinman}},\ }\bibfield  {title} {\bibinfo {title} {Nonlocal hermitian norm-conserving vanderbilt pseudopotential},\ }\href {https://doi.org/10.1103/PhysRevB.47.6728} {\bibfield  {journal} {\bibinfo  {journal} {Phys. Rev. B}\ }\textbf {\bibinfo {volume} {47}},\ \bibinfo {pages} {6728} (\bibinfo {year} {1993})}\BibitemShut {NoStop}%
\bibitem [{\citenamefont {Ozaki}(2003)}]{PhysRevB.67.155108}%
  \BibitemOpen
  \bibfield  {author} {\bibinfo {author} {\bibfnamefont {T.}~\bibnamefont {Ozaki}},\ }\bibfield  {title} {\bibinfo {title} {Variationally optimized atomic orbitals for large-scale electronic structures},\ }\href {https://doi.org/10.1103/PhysRevB.67.155108} {\bibfield  {journal} {\bibinfo  {journal} {Phys. Rev. B}\ }\textbf {\bibinfo {volume} {67}},\ \bibinfo {pages} {155108} (\bibinfo {year} {2003})}\BibitemShut {NoStop}%
\bibitem [{\citenamefont {Ozaki}()}]{OpenMX_basisset}%
  \BibitemOpen
  \bibfield  {author} {\bibinfo {author} {\bibfnamefont {T.}~\bibnamefont {Ozaki}},\ }\href@noop {} {\bibinfo {title} {Standard basis sets of openmx}},\ \bibinfo {note} {available at \url{https://www.openmx-square.org/openmx_man3.9/node27.html}}\BibitemShut {NoStop}%
\bibitem [{\citenamefont {Lejaeghere}\ \emph {et~al.}(2016)\citenamefont {Lejaeghere}, \citenamefont {Bihlmayer}, \citenamefont {Bj{\"o}rkman}, \citenamefont {Blaha}, \citenamefont {Bl{\"u}gel}, \citenamefont {Blum}, \citenamefont {Caliste}, \citenamefont {Castelli}, \citenamefont {Clark}, \citenamefont {Dal~Corso}, \citenamefont {de~Gironcoli}, \citenamefont {Deutsch}, \citenamefont {Dewhurst}, \citenamefont {Di~Marco}, \citenamefont {Draxl}, \citenamefont {Du{\l}ak}, \citenamefont {Eriksson}, \citenamefont {Flores-Livas}, \citenamefont {Garrity}, \citenamefont {Genovese}, \citenamefont {Giannozzi}, \citenamefont {Giantomassi}, \citenamefont {Goedecker}, \citenamefont {Gonze}, \citenamefont {Gr{\aa}n{\"a}s}, \citenamefont {Gross}, \citenamefont {Gulans}, \citenamefont {Gygi}, \citenamefont {Hamann}, \citenamefont {Hasnip}, \citenamefont {Holzwarth}, \citenamefont {Iu{\c s}an}, \citenamefont {Jochym}, \citenamefont {Jollet}, \citenamefont {Jones}, \citenamefont {Kresse}, \citenamefont {Koepernik},
  \citenamefont {K{\"u}{\c c}{\"u}kbenli}, \citenamefont {Kvashnin}, \citenamefont {Locht}, \citenamefont {Lubeck}, \citenamefont {Marsman}, \citenamefont {Marzari}, \citenamefont {Nitzsche}, \citenamefont {Nordstr{\"o}m}, \citenamefont {Ozaki}, \citenamefont {Paulatto}, \citenamefont {Pickard}, \citenamefont {Poelmans}, \citenamefont {Probert}, \citenamefont {Refson}, \citenamefont {Richter}, \citenamefont {Rignanese}, \citenamefont {Saha}, \citenamefont {Scheffler}, \citenamefont {Schlipf}, \citenamefont {Schwarz}, \citenamefont {Sharma}, \citenamefont {Tavazza}, \citenamefont {Thunstr{\"o}m}, \citenamefont {Tkatchenko}, \citenamefont {Torrent}, \citenamefont {Vanderbilt}, \citenamefont {van Setten}, \citenamefont {Van~Speybroeck}, \citenamefont {Wills}, \citenamefont {Yates}, \citenamefont {Zhang},\ and\ \citenamefont {Cottenier}}]{Lejaeghere2016-gh}%
  \BibitemOpen
  \bibfield  {author} {\bibinfo {author} {\bibfnamefont {K.}~\bibnamefont {Lejaeghere}}, \bibinfo {author} {\bibfnamefont {G.}~\bibnamefont {Bihlmayer}}, \bibinfo {author} {\bibfnamefont {T.}~\bibnamefont {Bj{\"o}rkman}}, \bibinfo {author} {\bibfnamefont {P.}~\bibnamefont {Blaha}}, \bibinfo {author} {\bibfnamefont {S.}~\bibnamefont {Bl{\"u}gel}}, \bibinfo {author} {\bibfnamefont {V.}~\bibnamefont {Blum}}, \bibinfo {author} {\bibfnamefont {D.}~\bibnamefont {Caliste}}, \bibinfo {author} {\bibfnamefont {I.~E.}\ \bibnamefont {Castelli}}, \bibinfo {author} {\bibfnamefont {S.~J.}\ \bibnamefont {Clark}}, \bibinfo {author} {\bibfnamefont {A.}~\bibnamefont {Dal~Corso}}, \bibinfo {author} {\bibfnamefont {S.}~\bibnamefont {de~Gironcoli}}, \bibinfo {author} {\bibfnamefont {T.}~\bibnamefont {Deutsch}}, \bibinfo {author} {\bibfnamefont {J.~K.}\ \bibnamefont {Dewhurst}}, \bibinfo {author} {\bibfnamefont {I.}~\bibnamefont {Di~Marco}}, \bibinfo {author} {\bibfnamefont {C.}~\bibnamefont {Draxl}}, \bibinfo {author}
  {\bibfnamefont {M.}~\bibnamefont {Du{\l}ak}}, \bibinfo {author} {\bibfnamefont {O.}~\bibnamefont {Eriksson}}, \bibinfo {author} {\bibfnamefont {J.~A.}\ \bibnamefont {Flores-Livas}}, \bibinfo {author} {\bibfnamefont {K.~F.}\ \bibnamefont {Garrity}}, \bibinfo {author} {\bibfnamefont {L.}~\bibnamefont {Genovese}}, \bibinfo {author} {\bibfnamefont {P.}~\bibnamefont {Giannozzi}}, \bibinfo {author} {\bibfnamefont {M.}~\bibnamefont {Giantomassi}}, \bibinfo {author} {\bibfnamefont {S.}~\bibnamefont {Goedecker}}, \bibinfo {author} {\bibfnamefont {X.}~\bibnamefont {Gonze}}, \bibinfo {author} {\bibfnamefont {O.}~\bibnamefont {Gr{\aa}n{\"a}s}}, \bibinfo {author} {\bibfnamefont {E.~K.~U.}\ \bibnamefont {Gross}}, \bibinfo {author} {\bibfnamefont {A.}~\bibnamefont {Gulans}}, \bibinfo {author} {\bibfnamefont {F.}~\bibnamefont {Gygi}}, \bibinfo {author} {\bibfnamefont {D.~R.}\ \bibnamefont {Hamann}}, \bibinfo {author} {\bibfnamefont {P.~J.}\ \bibnamefont {Hasnip}}, \bibinfo {author} {\bibfnamefont {N.~A.~W.}\ \bibnamefont
  {Holzwarth}}, \bibinfo {author} {\bibfnamefont {D.}~\bibnamefont {Iu{\c s}an}}, \bibinfo {author} {\bibfnamefont {D.~B.}\ \bibnamefont {Jochym}}, \bibinfo {author} {\bibfnamefont {F.}~\bibnamefont {Jollet}}, \bibinfo {author} {\bibfnamefont {D.}~\bibnamefont {Jones}}, \bibinfo {author} {\bibfnamefont {G.}~\bibnamefont {Kresse}}, \bibinfo {author} {\bibfnamefont {K.}~\bibnamefont {Koepernik}}, \bibinfo {author} {\bibfnamefont {E.}~\bibnamefont {K{\"u}{\c c}{\"u}kbenli}}, \bibinfo {author} {\bibfnamefont {Y.~O.}\ \bibnamefont {Kvashnin}}, \bibinfo {author} {\bibfnamefont {I.~L.~M.}\ \bibnamefont {Locht}}, \bibinfo {author} {\bibfnamefont {S.}~\bibnamefont {Lubeck}}, \bibinfo {author} {\bibfnamefont {M.}~\bibnamefont {Marsman}}, \bibinfo {author} {\bibfnamefont {N.}~\bibnamefont {Marzari}}, \bibinfo {author} {\bibfnamefont {U.}~\bibnamefont {Nitzsche}}, \bibinfo {author} {\bibfnamefont {L.}~\bibnamefont {Nordstr{\"o}m}}, \bibinfo {author} {\bibfnamefont {T.}~\bibnamefont {Ozaki}}, \bibinfo {author}
  {\bibfnamefont {L.}~\bibnamefont {Paulatto}}, \bibinfo {author} {\bibfnamefont {C.~J.}\ \bibnamefont {Pickard}}, \bibinfo {author} {\bibfnamefont {W.}~\bibnamefont {Poelmans}}, \bibinfo {author} {\bibfnamefont {M.~I.~J.}\ \bibnamefont {Probert}}, \bibinfo {author} {\bibfnamefont {K.}~\bibnamefont {Refson}}, \bibinfo {author} {\bibfnamefont {M.}~\bibnamefont {Richter}}, \bibinfo {author} {\bibfnamefont {G.-M.}\ \bibnamefont {Rignanese}}, \bibinfo {author} {\bibfnamefont {S.}~\bibnamefont {Saha}}, \bibinfo {author} {\bibfnamefont {M.}~\bibnamefont {Scheffler}}, \bibinfo {author} {\bibfnamefont {M.}~\bibnamefont {Schlipf}}, \bibinfo {author} {\bibfnamefont {K.}~\bibnamefont {Schwarz}}, \bibinfo {author} {\bibfnamefont {S.}~\bibnamefont {Sharma}}, \bibinfo {author} {\bibfnamefont {F.}~\bibnamefont {Tavazza}}, \bibinfo {author} {\bibfnamefont {P.}~\bibnamefont {Thunstr{\"o}m}}, \bibinfo {author} {\bibfnamefont {A.}~\bibnamefont {Tkatchenko}}, \bibinfo {author} {\bibfnamefont {M.}~\bibnamefont {Torrent}},
  \bibinfo {author} {\bibfnamefont {D.}~\bibnamefont {Vanderbilt}}, \bibinfo {author} {\bibfnamefont {M.~J.}\ \bibnamefont {van Setten}}, \bibinfo {author} {\bibfnamefont {V.}~\bibnamefont {Van~Speybroeck}}, \bibinfo {author} {\bibfnamefont {J.~M.}\ \bibnamefont {Wills}}, \bibinfo {author} {\bibfnamefont {J.~R.}\ \bibnamefont {Yates}}, \bibinfo {author} {\bibfnamefont {G.-X.}\ \bibnamefont {Zhang}},\ and\ \bibinfo {author} {\bibfnamefont {S.}~\bibnamefont {Cottenier}},\ }\bibfield  {title} {\bibinfo {title} {Reproducibility in density functional theory calculations of solids},\ }\href@noop {} {\bibfield  {journal} {\bibinfo  {journal} {Science}\ }\textbf {\bibinfo {volume} {351}},\ \bibinfo {pages} {aad3000} (\bibinfo {year} {2016})}\BibitemShut {NoStop}%
\bibitem [{\citenamefont {Ozaki}\ and\ \citenamefont {Kino}(2005)}]{PhysRevB.72.045121}%
  \BibitemOpen
  \bibfield  {author} {\bibinfo {author} {\bibfnamefont {T.}~\bibnamefont {Ozaki}}\ and\ \bibinfo {author} {\bibfnamefont {H.}~\bibnamefont {Kino}},\ }\bibfield  {title} {\bibinfo {title} {Efficient projector expansion for the ab initio lcao method},\ }\href {https://doi.org/10.1103/PhysRevB.72.045121} {\bibfield  {journal} {\bibinfo  {journal} {Phys. Rev. B}\ }\textbf {\bibinfo {volume} {72}},\ \bibinfo {pages} {045121} (\bibinfo {year} {2005})}\BibitemShut {NoStop}%
\bibitem [{\citenamefont {Duy}\ and\ \citenamefont {Ozaki}(2014)}]{DUY2014777}%
  \BibitemOpen
  \bibfield  {author} {\bibinfo {author} {\bibfnamefont {T.~V.~T.}\ \bibnamefont {Duy}}\ and\ \bibinfo {author} {\bibfnamefont {T.}~\bibnamefont {Ozaki}},\ }\bibfield  {title} {\bibinfo {title} {A three-dimensional domain decomposition method for large-scale dft electronic structure calculations},\ }\href {https://doi.org/https://doi.org/10.1016/j.cpc.2013.11.008} {\bibfield  {journal} {\bibinfo  {journal} {Computer Physics Communications}\ }\textbf {\bibinfo {volume} {185}},\ \bibinfo {pages} {777} (\bibinfo {year} {2014})}\BibitemShut {NoStop}%
\bibitem [{\citenamefont {Banerjee}\ \emph {et~al.}(1985)\citenamefont {Banerjee}, \citenamefont {Adams}, \citenamefont {Simons},\ and\ \citenamefont {Shepard}}]{doi:10.1021/j100247a015}%
  \BibitemOpen
  \bibfield  {author} {\bibinfo {author} {\bibfnamefont {A.}~\bibnamefont {Banerjee}}, \bibinfo {author} {\bibfnamefont {N.}~\bibnamefont {Adams}}, \bibinfo {author} {\bibfnamefont {J.}~\bibnamefont {Simons}},\ and\ \bibinfo {author} {\bibfnamefont {R.}~\bibnamefont {Shepard}},\ }\bibfield  {title} {\bibinfo {title} {Search for stationary points on surfaces},\ }\href {https://doi.org/10.1021/j100247a015} {\bibfield  {journal} {\bibinfo  {journal} {The Journal of Physical Chemistry}\ }\textbf {\bibinfo {volume} {89}},\ \bibinfo {pages} {52} (\bibinfo {year} {1985})}\BibitemShut {NoStop}%
\bibitem [{\citenamefont {Csaszar}\ and\ \citenamefont {Pulay}(1984)}]{CSASZAR198431}%
  \BibitemOpen
  \bibfield  {author} {\bibinfo {author} {\bibfnamefont {P.}~\bibnamefont {Csaszar}}\ and\ \bibinfo {author} {\bibfnamefont {P.}~\bibnamefont {Pulay}},\ }\bibfield  {title} {\bibinfo {title} {Geometry optimization by direct inversion in the iterative subspace},\ }\href {https://doi.org/https://doi.org/10.1016/S0022-2860(84)87198-7} {\bibfield  {journal} {\bibinfo  {journal} {Journal of Molecular Structure}\ }\textbf {\bibinfo {volume} {114}},\ \bibinfo {pages} {31} (\bibinfo {year} {1984})}\BibitemShut {NoStop}%
\bibitem [{\citenamefont {Broyden}(1970)}]{10.1093/imamat/6.1.76}%
  \BibitemOpen
  \bibfield  {author} {\bibinfo {author} {\bibfnamefont {C.~G.}\ \bibnamefont {Broyden}},\ }\bibfield  {title} {\bibinfo {title} {{The Convergence of a Class of Double-rank Minimization Algorithms 1. General Considerations}},\ }\href {https://doi.org/10.1093/imamat/6.1.76} {\bibfield  {journal} {\bibinfo  {journal} {IMA Journal of Applied Mathematics}\ }\textbf {\bibinfo {volume} {6}},\ \bibinfo {pages} {76} (\bibinfo {year} {1970})}\BibitemShut {NoStop}%
\bibitem [{\citenamefont {Fletcher}(1970)}]{10.1093/comjnl/13.3.317}%
  \BibitemOpen
  \bibfield  {author} {\bibinfo {author} {\bibfnamefont {R.}~\bibnamefont {Fletcher}},\ }\bibfield  {title} {\bibinfo {title} {{A new approach to variable metric algorithms}},\ }\href {https://doi.org/10.1093/comjnl/13.3.317} {\bibfield  {journal} {\bibinfo  {journal} {The Computer Journal}\ }\textbf {\bibinfo {volume} {13}},\ \bibinfo {pages} {317} (\bibinfo {year} {1970})}\BibitemShut {NoStop}%
\bibitem [{\citenamefont {Goldfarb}(1970)}]{10.2307/2004873}%
  \BibitemOpen
  \bibfield  {author} {\bibinfo {author} {\bibfnamefont {D.}~\bibnamefont {Goldfarb}},\ }\bibfield  {title} {\bibinfo {title} {A family of variable-metric methods derived by variational means},\ }\href {http://www.jstor.org/stable/2004873} {\bibfield  {journal} {\bibinfo  {journal} {Mathematics of Computation}\ }\textbf {\bibinfo {volume} {24}},\ \bibinfo {pages} {23} (\bibinfo {year} {1970})}\BibitemShut {NoStop}%
\bibitem [{\citenamefont {Shanno}(1970)}]{10.2307/2004840}%
  \BibitemOpen
  \bibfield  {author} {\bibinfo {author} {\bibfnamefont {D.~F.}\ \bibnamefont {Shanno}},\ }\bibfield  {title} {\bibinfo {title} {Conditioning of quasi-newton methods for function minimization},\ }\href {http://www.jstor.org/stable/2004840} {\bibfield  {journal} {\bibinfo  {journal} {Mathematics of Computation}\ }\textbf {\bibinfo {volume} {24}},\ \bibinfo {pages} {647} (\bibinfo {year} {1970})}\BibitemShut {NoStop}%
\bibitem [{\citenamefont {Fukuda}(2025{\natexlab{a}})}]{FLPQ}%
  \BibitemOpen
  \bibfield  {author} {\bibinfo {author} {\bibfnamefont {M.}~\bibnamefont {Fukuda}},\ }\href@noop {} {\bibinfo {title} {\textsc{FLPQ} package}} (\bibinfo {year} {2025}{\natexlab{a}}),\ \bibinfo {note} {\url{https://github.com/mfukudaQED/FLPQ}}\BibitemShut {NoStop}%
\bibitem [{\citenamefont {Senami}\ \emph {et~al.}()\citenamefont {Senami}, \citenamefont {Ichikawa}, \citenamefont {Tachibana}, \citenamefont {Fukuda},\ and\ \citenamefont {Soga}}]{QEDalpha}%
  \BibitemOpen
  \bibfield  {author} {\bibinfo {author} {\bibfnamefont {M.}~\bibnamefont {Senami}}, \bibinfo {author} {\bibfnamefont {K.}~\bibnamefont {Ichikawa}}, \bibinfo {author} {\bibfnamefont {A.}~\bibnamefont {Tachibana}}, \bibinfo {author} {\bibfnamefont {M.}~\bibnamefont {Fukuda}},\ and\ \bibinfo {author} {\bibfnamefont {K.}~\bibnamefont {Soga}},\ }\href@noop {} {\bibinfo {title} {\textsc{QEDalpha} package}},\ \bibinfo {note} {\url{https://github.com/mfukudaQED/QEDalpha}}\BibitemShut {NoStop}%
\bibitem [{\citenamefont {Fukuda}(2025{\natexlab{b}})}]{FLPQViewer}%
  \BibitemOpen
  \bibfield  {author} {\bibinfo {author} {\bibfnamefont {M.}~\bibnamefont {Fukuda}},\ }\href@noop {} {\bibinfo {title} {\textsc{FLPQViewer}}} (\bibinfo {year} {2025}{\natexlab{b}}),\ \bibinfo {note} {\url{https://github.com/mfukudaQED/FLPQViewer}}\BibitemShut {NoStop}%
\bibitem [{\citenamefont {Lee}\ and\ \citenamefont {Ozaki}(2019)}]{LEE2019192}%
  \BibitemOpen
  \bibfield  {author} {\bibinfo {author} {\bibfnamefont {Y.-T.}\ \bibnamefont {Lee}}\ and\ \bibinfo {author} {\bibfnamefont {T.}~\bibnamefont {Ozaki}},\ }\bibfield  {title} {\bibinfo {title} {\textsc{OpenMX Viewer}: A web-based crystalline and molecular graphical user interface program},\ }\href {https://doi.org/https://doi.org/10.1016/j.jmgm.2019.03.013} {\bibfield  {journal} {\bibinfo  {journal} {Journal of Molecular Graphics and Modelling}\ }\textbf {\bibinfo {volume} {89}},\ \bibinfo {pages} {192} (\bibinfo {year} {2019})}\BibitemShut {NoStop}%
\bibitem [{\citenamefont {Kawamura}(2019)}]{KAWAMURA2019197}%
  \BibitemOpen
  \bibfield  {author} {\bibinfo {author} {\bibfnamefont {M.}~\bibnamefont {Kawamura}},\ }\bibfield  {title} {\bibinfo {title} {\textsc{FermiSurfer}: Fermi-surface viewer providing multiple representation schemes},\ }\href {https://doi.org/https://doi.org/10.1016/j.cpc.2019.01.017} {\bibfield  {journal} {\bibinfo  {journal} {Computer Physics Communications}\ }\textbf {\bibinfo {volume} {239}},\ \bibinfo {pages} {197} (\bibinfo {year} {2019})}\BibitemShut {NoStop}%
\bibitem [{\citenamefont {Gross}\ \emph {et~al.}(2018)\citenamefont {Gross}, \citenamefont {Schuler}, \citenamefont {Pavliček}, \citenamefont {Fatayer}, \citenamefont {Majzik}, \citenamefont {Moll}, \citenamefont {Peña},\ and\ \citenamefont {Meyer}}]{https://doi.org/10.1002/anie.201703509}%
  \BibitemOpen
  \bibfield  {author} {\bibinfo {author} {\bibfnamefont {L.}~\bibnamefont {Gross}}, \bibinfo {author} {\bibfnamefont {B.}~\bibnamefont {Schuler}}, \bibinfo {author} {\bibfnamefont {N.}~\bibnamefont {Pavliček}}, \bibinfo {author} {\bibfnamefont {S.}~\bibnamefont {Fatayer}}, \bibinfo {author} {\bibfnamefont {Z.}~\bibnamefont {Majzik}}, \bibinfo {author} {\bibfnamefont {N.}~\bibnamefont {Moll}}, \bibinfo {author} {\bibfnamefont {D.}~\bibnamefont {Peña}},\ and\ \bibinfo {author} {\bibfnamefont {G.}~\bibnamefont {Meyer}},\ }\bibfield  {title} {\bibinfo {title} {Atomic force microscopy for molecular structure elucidation},\ }\href {https://doi.org/https://doi.org/10.1002/anie.201703509} {\bibfield  {journal} {\bibinfo  {journal} {Angewandte Chemie International Edition}\ }\textbf {\bibinfo {volume} {57}},\ \bibinfo {pages} {3888} (\bibinfo {year} {2018})}\BibitemShut {NoStop}%
\bibitem [{\citenamefont {Carey}\ and\ \citenamefont {Sundberg}(2007)}]{carey2007advanced}%
  \BibitemOpen
  \bibfield  {author} {\bibinfo {author} {\bibfnamefont {F.~A.}\ \bibnamefont {Carey}}\ and\ \bibinfo {author} {\bibfnamefont {R.~J.}\ \bibnamefont {Sundberg}},\ }\href {https://doi.org/10.1007/978-0-387-44899-2} {\emph {\bibinfo {title} {Advanced Organic Chemistry: Part A: Structure and Mechanisms}}},\ \bibinfo {edition} {5th}\ ed.\ (\bibinfo  {publisher} {Springer},\ \bibinfo {year} {2007})\BibitemShut {NoStop}%
\bibitem [{\citenamefont {Hirao}\ and\ \citenamefont {Ohwada}(2003)}]{doi:10.1021/jp027330l}%
  \BibitemOpen
  \bibfield  {author} {\bibinfo {author} {\bibfnamefont {H.}~\bibnamefont {Hirao}}\ and\ \bibinfo {author} {\bibfnamefont {T.}~\bibnamefont {Ohwada}},\ }\bibfield  {title} {\bibinfo {title} {Theoretical study of reactivities in electrophilic aromatic substitution reactions: Reactive hybrid orbital analysis},\ }\href {https://doi.org/10.1021/jp027330l} {\bibfield  {journal} {\bibinfo  {journal} {The Journal of Physical Chemistry A}\ }\textbf {\bibinfo {volume} {107}},\ \bibinfo {pages} {2875} (\bibinfo {year} {2003})}\BibitemShut {NoStop}%
\bibitem [{\citenamefont {Bader}(1990)}]{10.1093/oso/9780198551683.001.0001}%
  \BibitemOpen
  \bibfield  {author} {\bibinfo {author} {\bibfnamefont {R.~F.~W.}\ \bibnamefont {Bader}},\ }\href {https://doi.org/10.1093/oso/9780198551683.001.0001} {\emph {\bibinfo {title} {Atoms in Molecules: A Quantum Theory}}}\ (\bibinfo  {publisher} {Oxford University Press},\ \bibinfo {year} {1990})\BibitemShut {NoStop}%
\bibitem [{\citenamefont {Parr}\ and\ \citenamefont {Weitao}(1995)}]{10.1093/oso/9780195092769.001.0001}%
  \BibitemOpen
  \bibfield  {author} {\bibinfo {author} {\bibfnamefont {R.~G.}\ \bibnamefont {Parr}}\ and\ \bibinfo {author} {\bibfnamefont {Y.}~\bibnamefont {Weitao}},\ }\href {https://doi.org/10.1093/oso/9780195092769.001.0001} {\emph {\bibinfo {title} {Density-Functional Theory of Atoms and Molecules}}}\ (\bibinfo  {publisher} {Oxford University Press},\ \bibinfo {year} {1995})\BibitemShut {NoStop}%
\end{thebibliography}%

\clearpage


\setcounter{equation}{0}
\setcounter{figure}{0}
\setcounter{table}{0}
\setcounter{page}{1}
\setcounter{section}{0}
\makeatletter
\renewcommand{\theequation}{S\arabic{equation}}
\renewcommand{\thefigure}{S\arabic{figure}}
\renewcommand{\thetable}{S\arabic{table}}
\renewcommand{\thesection}{S\arabic{section}}
\renewcommand{\thepage}{S\arabic{page}}
\renewcommand{\appendixname}{}

\def\PATHFIG_supp{./fig_sup}

\onecolumngrid

\section*{Supplemental Information}

\section{Electronic transfer for ionic bonding} \label{sec:ionic}
Suppose that there are two isolated states A and B.
\begin{align}
\Delta E = \left(\mu_A^0 - \mu_B^0 \right) \Delta N+ \left(\eta_A+\eta_B \right)(\Delta N)^2,
\end{align}
\begin{align}
 \Delta N &\equiv N_B^0 - N_B = N_A - N_A^0, \\
 &= \frac{\mu_B^0 - \mu_A^0}{2(\eta_A + \eta_B)},
\end{align}
\begin{align}
\Delta E = -\frac{\left(\mu_B^0 - \mu_A^0 \right)^2}{4(\eta_A+\eta_B)}.
\end{align}

\section{Energy decomposition and virial theorem} \label{sec:virial}
Total energy of the Rigged QED system (electron, photon, and nuclei) is defined
as the integral of the 0-0 component of the energy momentum tensor $T^{\mu \nu}$,
which consists of three contributions from electron field, nucleus field, and photon field.
\begin{align}
\hat{T}^{\mu \nu} = \hat{T}_e^{\mu \nu} + \sum_a \hat{T}_a^{\mu \nu} + \hat{T}_{\rm EM}^{\mu \nu},
\end{align}
If the system is in an equilibrium stationary state,
the virial theorem can be applied, and we obtain the following relation.
\begin{align}
E_{\rm Rigged\ QED} &\equiv \left\langle \int {\hat{T}}_0^{\ 0} d^3\vec{r} \right\rangle, \\
&= \left\langle \int \hat{T}_{e \mu}{}^{\mu} d^3\vec{r} \right\rangle
+ \sum_a \left\langle \int  \hat{T}_{a \mu}{}^{\mu} d^3\vec{r} \right\rangle.
\end{align}
In the virial theorem, we used the conservation law of the energy momentum tensor, $\partial_\nu \hat{T}^{\nu\mu} = 0$, and we assumed at infinity $\hat{T}^{\mu \nu} = 0$. 
In other words, $\left\langle \int \hat{T}_i^{\ i} d^3 \vec{r} \right\rangle = 0$.

The electron part of the above energy expression can be expanded as follows. 
\begin{align}
E_{e \rm Rigged\ QED}
&\equiv \left\langle \int \hat{T}_{e \mu}{}^{\mu} d^3\vec{r} \right\rangle,\\
&=\left\langle \int \left[ \frac{1}{2} \left( \hat{M}_e + h.c. \right) + \frac{1}{2} \left( i\hbar c \hat{\bar{\psi}} \gamma^k \hat{D}_{ek} \hat{\psi} + h.c. \right)  \right] d^3\vec{r} \right\rangle,\\
&=\left\langle \int \left[ m_ec^2 \hat{\bar{\psi}} \hat{\psi}  \right] d^3\vec{r} \right\rangle,
\end{align}
where $\bar{\psi} \equiv \hat{\psi^\dagger \gamma^0}$.
The matrices $\gamma^0$ and $\gamma^k$ are the 4 × 4 gamma matrices.
The four-component $\psi$ can be expressed by two-component $\psi^L$ and $\psi^S$.
The $\psi^S$ can be expressed by $\psi^L$ in the time-independent stationary state as follows~\cite{Fukuda2016_LPQ}.
\begin{align}
\hat{\psi} = 
\begin{pmatrix}
\hat{\psi}^L \\
\hat{\psi}^S
\end{pmatrix}, 
\ \ \ \hat{\psi}^S \approx \frac{-i \hbar \bar{K}}{2m_e c} \sigma^m \partial_m \hat{\psi}^L,
\ \ \ \bar{K} = \frac{2m_e c^2}{2m_e c^2 + E_{e \rm Rigged\ QED} -Z_e e A_0},
\end{align}
where $A_0$ is the scalar potential.
The above relation is obtained by the Dirac equation.
Under the non-relativistic limit, $\bar{K} \approx 1$.
For simplicity, we ignored the vector potential. 

To obtain the energy in the non-relativistic limit,
we subtract the mass energy from the energy.
\begin{align}
\nonumber
& E_{e \rm Rigged\ QED} - \left\langle \int m_ec^2 \hat{\psi}^\dagger \hat{\psi} d^3 \vec{r}\right\rangle,\\
\nonumber
&= \left\langle \int \left[ m_ec^2 \hat{\bar{\psi}} \hat{\psi}  \right] d^3\vec{r} \right\rangle
- \left\langle \int m_ec^2 \hat{\psi}^\dagger \hat{\psi} d^3 \vec{r} \right\rangle,\\
\nonumber
&= \left\langle \int \left[ m_ec^2 \left( \hat{\psi}^{L\dagger} \hat{\psi}^{L} -  \hat{\psi}^{S\dagger} \hat{\psi}^S \right)  \right] d^3\vec{r} \right\rangle - \left\langle \int \left[ m_ec^2 \left( \hat{\psi}^{L\dagger} \hat{\psi}^{L} + \hat{\psi}^{S\dagger} \hat{\psi}^S \right)  \right] d^3\vec{r} \right\rangle,\\
\nonumber
&= \left\langle \int \left[ -2m_ec^2 \hat{\psi}^{S\dagger} \hat{\psi}^S  \right] d^3\vec{r} \right\rangle,\\
&\approx E_e^{NR},
\end{align}
\begin{align}
\nonumber
E_e^{NR} &= \left\langle \int \left[ -2m_ec^2 \left( \frac{\hbar^2 \bar{K}^2}{4m_e^2 c^2} \left(  \partial_l \hat{\psi}^L \right)^\dagger \sigma^l \sigma^m \partial_m \hat{\psi}^L \right)  \right] d^3\vec{r}
\right\rangle,\\
\nonumber
&= \left\langle \int \left[ -2m_ec^2 \left( \frac{\hbar^2 \bar{K}^2}{4m_e^2 c^2} \left( \left(  \partial_l \hat{\psi}^L \right)^\dagger \partial_l \hat{\psi}^L + i \varepsilon_{lmn} \left(\partial_l \hat{\psi}^L \right)^\dagger \sigma^n \partial_m \hat{\psi}^L \right) \right)  \right] d^3\vec{r}
\right\rangle,\\
\nonumber
&= \left\langle \int \left[ -2m_ec^2 \left(\frac{\hbar^2 \bar{K}^2}{4m_e^2 c^2} \left(   \partial_l \hat{\psi}^L \right)^\dagger \partial_l \hat{\psi}^L \right)  \right] d^3\vec{r}
\right\rangle,\\
\nonumber
&= \left\langle \int \left[ -2m_ec^2 \left(- \frac{\hbar^2 \bar{K}^2}{8m_e^2 c^2} \left( \hat{\psi}^{L\dagger} \partial_l \partial_l \hat{\psi}^L + \left( \partial_l \partial_l \hat{\psi}^L \right)^\dagger \hat{\psi}^L \right) \right)  \right] d^3\vec{r}
\right\rangle,\\
\nonumber
&= \left\langle \int \left[ \frac{\hbar^2 \bar{K}^2}{4m_e} \left( \left( \hat{\psi}^{L\dagger} \partial_l \partial_l \hat{\psi}^L + \left( \partial_l \partial_l \hat{\psi}^L \right)^\dagger \hat{\psi}^L \right) \right)  \right] d^3\vec{r}
\right\rangle,\\
\nonumber
&= \left\langle \int \left[ -\frac{1}{2}  \left(- \frac{\hbar^2 \bar{K}^2}{2m_e} \hat{\psi}^{L\dagger} \partial_l \partial_l \hat{\psi}^L + h.c. \right)  \right] d^3\vec{r} 
\right\rangle,\\
&= - \langle \hat{T}_e \rangle .
\end{align}
Here, the kinetic energy density operator is defined as
\begin{align}
    \hat{T}_e &\equiv - \frac{\hbar^2}{4m_e} \hat{\psi}^{\dagger} \hat{D}_{el} \hat{D}_{el} \hat{\psi} + h.c.,\\
              &\approx - \frac{\hbar^2 \bar{K}^2}{4m_e} \hat{\psi}^{L\dagger} \partial_l \partial_l \hat{\psi}^L + h.c.
\end{align}
Similarly, the energy of the nuclei part is written by the kinetic energy of the nuclei too.
Therefore,
\begin{align}
E_{\rm Rigged\ QED} &= - \langle \hat{T}_e \rangle - \sum_a \langle \hat{T}_a \rangle.
\end{align}
Under the Born-Oppenheimer approximation,
\begin{align}
E_{\rm Rigged\ QED} &\approx - \langle \hat{T}_e \rangle. 
\end{align}
Furthermore, the kinetic energy of the electron can be expanded
by minus half of the integral over the trace of the electronic stress tensor density.
Therefore,
\begin{align}
E_{\rm Rigged\ QED} &\approx - \langle \hat{T}_e \rangle = \frac{1}{2}\int d^3 \vec{r} \left\langle  \sum_{k=1}^3 \hat{\tau}_e^{kk} (\vec{r}) \right\rangle. 
\end{align}
This expression gives an energy decomposition scheme using electronic stress tensor density.
The regional energy density operator constructed from the electronic stress tensor density operator is defined as follows.
\begin{align}
\hat{\varepsilon}_{\tau} (\vec{r}) &\equiv \frac{1}{2} \sum_{k=1}^3  \hat{\tau}^{Skk} (\vec{r}\,),\\
E_{\rm Rigged\ QED} &\approx \int d^3 \vec{r} \left\langle \hat{\varepsilon}_{\tau} (\vec{r}\,)\right\rangle .
\end{align}
This means that the energy density is decomposed into the sum of the eigenvalue of electronic stress tensor density.

In the non-relativistic limit,
\begin{align}
\nonumber
\hat{\varepsilon}_{\tau} (\vec{r}\,)
\approx 
\hat{\varepsilon}^{\rm NR}_{\tau} (\vec{r}\,) &= \frac{\hbar^2}{8m} \sum_{k=1}^3 \left[ \hat{\psi}^{L\dagger} \partial_k \partial_k \hat{\psi}^{L\dagger} - \left(\partial_k \hat{\psi}^L \right)^\dagger \partial_k \hat{\psi}^L + h.c. \right],\\
\nonumber
&= \frac{\hbar^2}{8m} \sum_{k=1}^3 \left[ 2\hat{\psi}^{L\dagger} \partial_k \partial_k \hat{\psi}^L - \partial_k \left(\hat{\psi}^{L\dagger} \partial_k \hat{\psi}^L \right) + h.c. \right],\\
&= \frac{\hbar^2}{8m} \sum_{k=1}^3 \left[ 2\hat{\psi}^{L\dagger} \partial_k \partial_k \hat{\psi}^L - \frac{1}{2} \partial_k \partial_k \left( \hat{\psi}^{L\dagger} \hat{\psi}^L \right) + h.c. \right] .
\end{align}
For simplicity, we ignored the vector potential. 

At the electronic interface, where the kinetic energy density equals zero,
\begin{align}
\hat{\varepsilon}^{\rm NR}_{\tau} ({\rm surface}) 
\nonumber
&= - \frac{\hbar^2}{8m} \sum_{k=1}^3 \left[ \partial_k \partial_k \left( \hat{\psi}^{L\dagger} \hat{\psi}^L \right) \right],\\
&= - \frac{\hbar^2}{8m} \nabla^2 \hat{n}.
\end{align}
In AIM theory, $ \langle \nabla^2 \hat{n} \rangle$ is used as a measure of localization in the bonding region.

\section{Local mechanical picture of covalent bonding} \label{sec:fig_Covalent_bonding}
\begin{figure*}[htb]
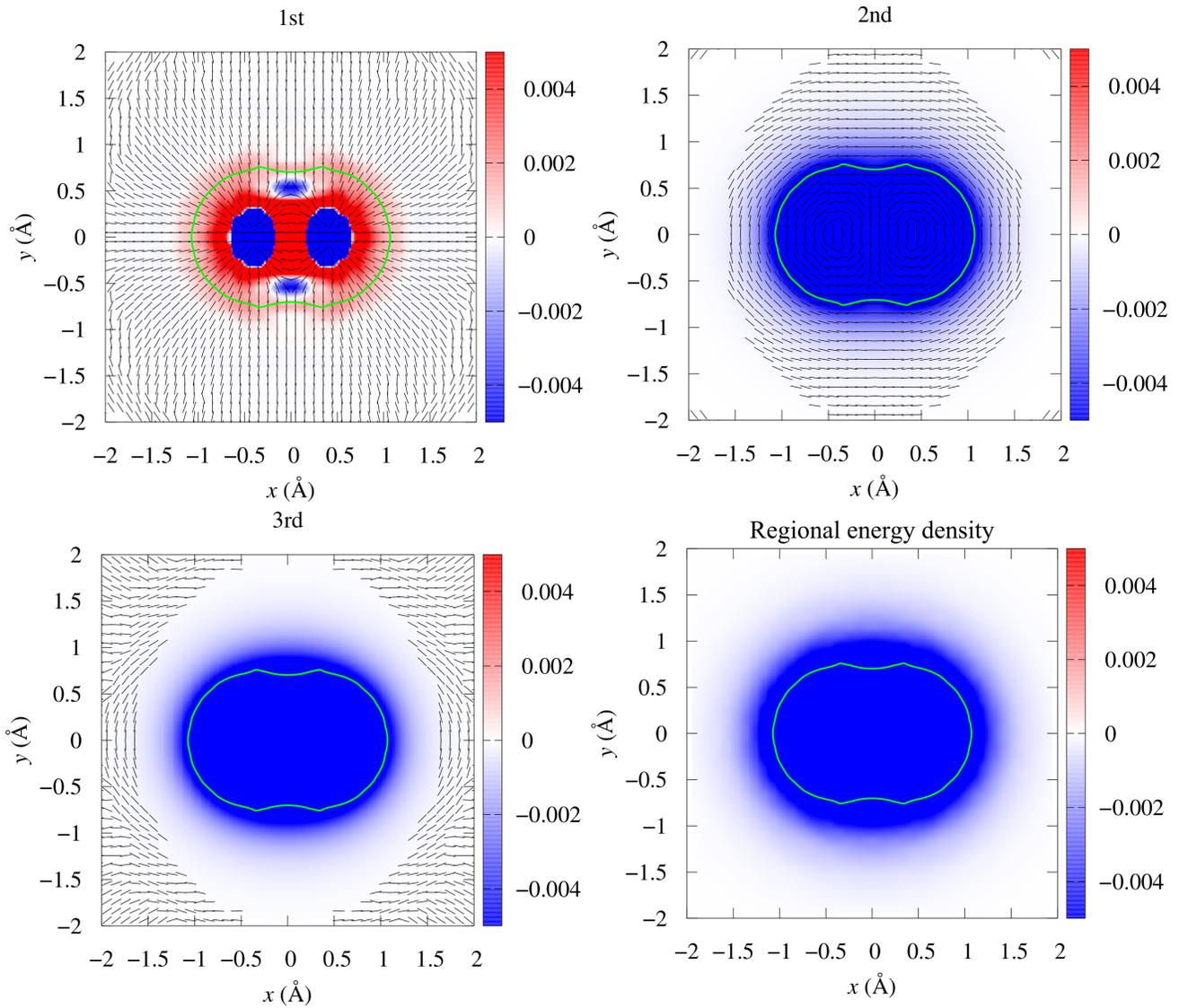

    \centering
  \begin{minipage}[b]{0.48\linewidth}
    \centering
    \includegraphics[width=\linewidth]{\PATHFIG_supp/H2_stress_1st.png}
  \end{minipage}
  \begin{minipage}[b]{0.48\linewidth}
    \includegraphics[width=\linewidth]{\PATHFIG_supp/H2_stress_2nd.png}
  \end{minipage}
  \\
   \begin{minipage}[b]{0.48\linewidth}
    \centering
    \includegraphics[width=\linewidth]{\PATHFIG_supp/H2_stress_3rd.png}
  \end{minipage}
  \begin{minipage}[b]{0.48\linewidth}
    \includegraphics[width=\linewidth]{\PATHFIG_supp/H2_energy_density.png}
  \end{minipage} 
        \caption{The principal electronic stress tensor density and energy density of H$_2$. The eigenvalue of the regional energy density (color map) and corresponding eigenvector (black rods). The lines in green represent the electronic interface (material surface).   }\label{fig:stress_H2}
\end{figure*}

\begin{figure*}[htb]
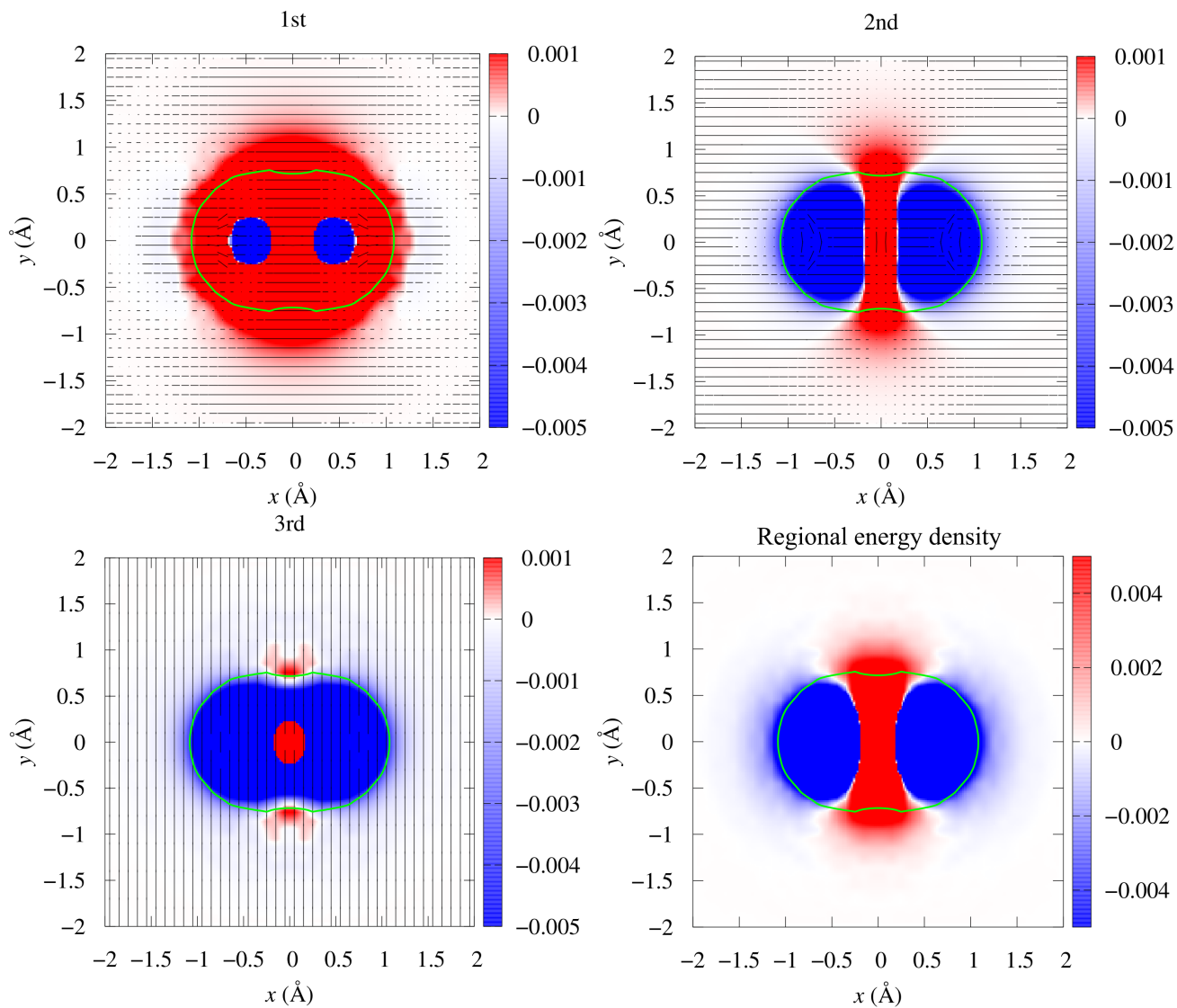

    \centering
  \begin{minipage}[b]{0.48\linewidth}
    \centering
    \includegraphics[width=\linewidth]{\PATHFIG_supp/diff_H2_stress_1st.png}
  \end{minipage}
  \begin{minipage}[b]{0.48\linewidth}
    \includegraphics[width=\linewidth]{\PATHFIG_supp/diff_H2_stress_2nd.png}
  \end{minipage}
  \\
   \begin{minipage}[b]{0.48\linewidth}
    \centering
    \includegraphics[width=\linewidth]{\PATHFIG_supp/diff_H2_stress_3rd.png}
  \end{minipage}
  \begin{minipage}[b]{0.48\linewidth}
    \includegraphics[width=\linewidth]{\PATHFIG_supp/diff_H2_energy_density.png}
  \end{minipage} 
        \caption{The principal electronic stress tensor density and energy density of H$_2$ $-$ 2H. The eigenvalue of the electronic stress tensor density (color map) and corresponding eigenvector (black rods). 
        The figures show the difference between the physical quantity of a hydrogen molecule and that of two hydrogen atoms.
        The lines in green represent the electronic interface (material surface)}\label{fig:diff_stress_H2}
\end{figure*}

The local mechanical picture of covalent bonding for the H$_2$ molecule is visualized in Figs.~\ref{fig:stress_H2} and ~\ref{fig:diff_stress_H2}
by the eigenvalues and eigenvectors of the electronic stress tensor density and the regional energy density.

\clearpage

\end{document}